\documentclass[english,aip,jcp,reprint]{revtex4-1}
\usepackage[T1]{fontenc}
\usepackage[utf8]{inputenc}
\usepackage{refstyle}
\usepackage{booktabs}
\usepackage{amsmath}
\usepackage{amssymb}
\usepackage{graphicx}
\usepackage{color}

\makeatletter

\AtBeginDocument{\providecommand\eqref[1]{\ref{eq:#1}}}
\AtBeginDocument{\providecommand\tabref[1]{\ref{tab:#1}}}
\AtBeginDocument{\providecommand\figref[1]{\ref{fig:#1}}}
\providecommand{\tabularnewline}{\\}
\RS@ifundefined{subsecref}
  {\newref{subsec}{name = \RSsectxt}}
  {}
\RS@ifundefined{thmref}
  {\def\RSthmtxt{theorem~}\newref{thm}{name = \RSthmtxt}}
  {}
\RS@ifundefined{lemref}
  {\def\RSlemtxt{lemma~}\newref{lem}{name = \RSlemtxt}}
  {}

\usepackage{braket}
\usepackage{simplewick}
\usepackage[version=3]{mhchem}

\@ifundefined{showcaptionsetup}{}{%
 \PassOptionsToPackage{caption=false}{subfig}}
\usepackage{subfig}
\makeatother

\usepackage{babel}
\begin{document}

\title{Cluster decomposition of full configuration interaction wave functions:
a tool for chemical interpretation of systems with strong correlation}

\author{Susi Lehtola}

\affiliation{Chemical Sciences Division, Lawrence Berkeley National Laboratory,
Berkeley, California 94720, United States}
\email{susi.lehtola@alumni.helsinki.fi}

\author{Norm M. Tubman}

\affiliation{University of California, Berkeley, California 94720, United States}

\author{K. Birgitta Whaley}

\affiliation{University of California, Berkeley, California 94720, United States}

\author{Martin Head-Gordon}

\affiliation{Chemical Sciences Division, Lawrence Berkeley National Laboratory,
Berkeley, California 94720, United States}

\affiliation{University of California, Berkeley, California 94720, United States}

\begin{abstract}
Approximate full configuration interaction (FCI) calculations have
recently become tractable for systems of unforeseen size thanks to
stochastic and adaptive approximations to the exponentially scaling
FCI problem.  The result of an FCI calculation is a weighted set of
electronic configurations, which can also be expressed in terms of
excitations from a reference configuration.  The excitation amplitudes
contain information on the complexity of the electronic wave function,
but this information is contaminated by contributions from
disconnected excitations, i.e. those excitations that are just
products of independent lower-level excitations. The unwanted
contributions can be removed via a cluster decomposition procedure,
making it possible to examine the importance of connected excitations
in complicated multireference molecules which are outside the reach of
conventional algorithms. We present an implementation of the cluster
decomposition analysis and apply it to both true FCI wave functions,
as well as wave functions generated from the adaptive sampling CI
(ASCI) algorithm. The cluster decomposition is useful for interpreting
calculations in chemical studies, as a diagnostic for the convergence
of various excitation manifolds, as well as as a guidepost for
polynomially scaling electronic structure models. Applications are
presented for (i) the double dissociation of water, (ii) the carbon
dimer, (iii) the $\pi$ space of polyacenes, as well as (iv) the
chromium dimer. While the cluster amplitudes exhibit rapid decay with
increasing rank for the first three systems, even connected octuple
excitations still appear important in \ce{Cr2}, suggesting that
spin-restricted single-reference coupled-cluster approaches may not be
tractable for some problems in transition metal chemistry.
\end{abstract}
\maketitle
\global\long\def\ERI#1#2{(#1|#2)}
\global\long\def\bra#1{\Bra{#1}}
\global\long\def\ket#1{\Ket{#1}}
\global\long\def\braket#1{\Braket{#1}}

\newcommand{\citeref}[1]{reference \citenum{#1}}

\section{Introduction\label{sec:Introduction}}

The main problem of quantum chemistry is to solve the Schrödinger
equation
\begin{equation}
  \hat{H} \ket{\Psi} = E \ket{\Psi} \label{eq:schr}
\end{equation}
for the many-electron wave function $\ket{\Psi}$. The wave function
can be expanded in terms of a set of single-particle states
$\{\psi_i\}$ (commonly referred to as orbitals) through the resolution
of the identity $\sum_i \ket{\psi_i} \bra{\psi_i} = 1 $ as
\begin{equation}
  \ket{\Psi} = c_{n_1,\dots,n_i,\dots,n_j,\dots} \ket{\psi_{n_1}} \cdots \ket{\psi_{n_i}} \cdots \ket{\psi_{n_j}} \cdots \label{eq:orbexp}
\end{equation}
Because the wave function must satisfy Fermi statistics $\Psi({\bf
  x}_1,\dots,{\bf x}_i,\dots,{\bf x}_j,\dots) = -\Psi({\bf
  x}_1,\dots,{\bf x}_j,\dots,{\bf x}_i,\dots)$, it is seen that the
tensor $c$ must be completely antisymmetric. However, its form is
otherwise unknown \emph{a priori}.

In order to motivate the topic of the present manuscript -- the
cluster decomposition -- we begin by describing two well-known
parametrizations for the many-electron wave function $\ket{\Psi}$:
configuration interaction (CI) theory and coupled-cluster (CC)
theory. CI and CC are equivalent at the limit of exactness: when all
possible electronic excitations are included in the models, yielding
the full CI (FCI) and full CC (FCC) methods, the models only differ in
the way the wave function is parametrized. The cluster decomposition
is a way of converting FCI wave functions to FCC wave functions (and
vice versa), thus allowing for each method to be used where it is
technically most favorable, all the while enabling one to elucidate
the structure of the resulting many-electron wave function.

\subsection{Configuration interaction\label{sub:ConfigurationInteraction}}

In configuration interaction theory, \eqref{orbexp} is recast into an
equivalent form based on Slater determinants $\ket{\Phi_k}$
\begin{equation}
  \ket{\Psi} = \sum_{k=0}^{N_\text{dets}-1} C_k \ket{\Phi_k} \label{eq:slatexp}
\end{equation}
which builds in the necessary spin statistics and is accessible to
numerical implementation. A Slater determinant is defined by the
orbitals that the electrons occupy in it, which is why it is often
referred to as an electron configuration. The expansion coefficients
$C_k$ (as well as the orbitals $\psi_n$ in case only a subset of
configurations are included -- such as $N_\text{dets}=1$ in the
Hartree--Fock model) can be solved by minimizing the expectation value
$\braket{\Psi | \hat{H} | \Psi}$. This results in a linear eigenvalue
equation for the expansion coefficients $C_k$.

Thanks to its simplicity and its fulfillment of a variational theorem,
CI theory is the traditional approach for solving the Schrödinger
equation in a given one-electron basis set. The exact solution to the
Schrödinger equation can be obtained by including all the electronic
configurations in the problem, via the use of the FCI method. However,
the dimension of the FCI wave function -- the number of possible
Slater determinants for $n$ electrons in $N$ orbitals, denoted as
($n$e,$N$o), scales as (ignoring any symmetries)
\begin{align}
\dim\mathcal{H}\approx &
\left(N!/\left[\left(N-n/2\right)!\left(n/2\right)!\right]\right)^{2},\label{eq:ndets}
\end{align}
highlighting the difficulty of the exact solution to the Schrödinger
equation. As can be seen from \eqref{ndets}, the dimension of the wave
function increases exponentially with increasing system size, and this
exponential wall\cite{Kohn1999} is not affected by the massive
increase of computational power given by Moore's
law\cite{Moore1998}. Millions of determinants \textendash{} e.g. the
(13e,13o) problem \textendash{} could be handled in the early
1980s,\cite{Saxe1981} while billion-determinant FCI calculations
\textendash{} corresponding to (18e,18o) \textendash{} have been
around since the late 1980s.\cite{Knowles1989, Knowles1989b,
  Olsen1990} The best calculations to date have only included tens of
billions of determinants,\cite{Rossi1999, Thogersen2004a, Gan2005,
  Gan2006} i.e.  (20e,20o), with problem sizes close to a trillion
determinants i.e.  (24e,24o) becoming feasible in the near future on
massively parallel architectures.\cite{Vogiatzis2017} Due to the steep
scaling of FCI, many different models have been proposed over the
years to selectively capture correlation effects in the wave function,
which will be discussed next.

If good orbitals are used for the problem, the expansion of
\eqref{slatexp} for a ground state is often dominated by the Aufbau
configuration $\Phi_0$.  In this case, it is appealing to rewrite the
wave function in terms of excitations from the reference configuration
as
\begin{equation}
  \ket{\Psi} = \mathcal{N} \left( 1 + \hat{C}_1 + \hat{C}_2 + \dots \right)
  \ket{\Phi_0}, \label{eq:ciexp}
\end{equation}
where $\mathcal{N}$ is a normalization constant, and $\hat{C}_n$ is an
$n$-fold excitation operator. Typically, one sets $\mathcal{N}=1$ to
simplify manipulations -- this is known as intermediate normalization
-- requiring that the normalization be restored at the end in any
calculations. A natural suggestion to circumvent the exponential
scaling of FCI with system size would be to impose restrictions on the
allowed configuration space, such as only considering single and
double excitations from the reference by setting $\hat{C}_n=0$ for
$n>2$ in \eqref{ciexp}. This yields the configuration interaction
singles and doubles (CISD) method. Unfortunately, while it is still
variational, truncated CI is not size consistent. CISD is exact for a
single helium atom, but when applied to \ce{He2} its performance is
degraded, as the triple and quadruple excitations necessary for
exactness for the four-electron system are excluded \textendash{} even
if the two atoms are infinitely far apart.

Alternatively, the cost may be brought down by reducing the number of
electrons and orbitals considered in the FCI problem. This is the
approach pursued in multiconfigurational (MC) self-consistent field
(SCF) theory in general, and in its most popular variant, the complete
active space (CAS) SCF method,\cite{Roos1980, Roos1980a} where the FCI
problem is only solved within an orbital active space. Developments on
the CASSCF method, such as the generalized active space (GAS) SCF
method,\cite{Olsen1983, Fleig2001, Ma2011} further tailor the number
of configurations considered in the wave function by the introduction
of occupancy restrictions on the active orbitals.\cite{Walch1983,
  Olsen1988, Panin1996, Panin1996a, Fleig2001, Nakano2000, Ivanic2003,
  Ma2011} It is also possible to reduce the number of configurations
in the diagonalization procedure by expressing a part of them only
perturbatively.\cite{Lowdin1951, LiManni2011, LiManni2013}

Still, a FCI or CASSCF wave function typically contains many
insignificant configurations: chemical accuracy ($10^{-3}E_{h}$) is
achievable by retaining only a small fraction of the
configurations.\cite{Ivanic2001, Ivanic2002, Bytautas2003,
  Bytautas2009} With a smart way of choosing the configurations
$\ket{\Phi_k}$ that are included in the expansion (\eqref{slatexp}),
accurate solutions may still be achieved for systems that would be too
difficult to solve exactly. Some of us have recently developed a fast
and efficient CI approach called adaptive sampling CI
(ASCI),\cite{Tubman2016} in which only the important configurations
are considered. There has been much recent interest in the ASCI
approach due to its demonstrated ability to treat strongly correlated
systems.  The inspiration for the ASCI technique is in selected CI
methods\cite{Bender1969, Langhoff1973, Huron1973, Buenker1974,
  Buenker1978, Evangelisti1983, Cimiraglia1987, Illas1991,
  Harrison1991, Daudey1993, Neese2003, Roth2009} and FCI quantum Monte
Carlo methods\cite{Booth2009, Booth2010, Booth2011, BenAmor2011,
  Shepherd2012, Shepherd2012a, Shepherd2012b, Daday2012, Giner2013,
  Evangelista2014, Liu2014, Thomas2015a, Giner2015}. Based on these
ideas, a variety of improvements and other deterministic-adaptive
approaches have also been recently suggested.\cite{Holmes2016a,
  Schriber2016, Liu2016} In addition, several incremental schemes that
do not yield a wave function have also been recently proposed for the
estimation of FCI energies.\cite{Zimmerman2017, *Zimmerman2017a,
  *Zimmerman2017b, Eriksen2017}

In contrast to both the CAS approach, where the smallness of the
feasible active space poses significant limitations, and to the RAS
and GAS approaches, which allow for slightly larger active spaces at
the cost of having to specify the structure of the wave function --
that is often far from trivial -- the adaptive approaches can handle
large active spaces with minimal user intervention. For instance, the
ASCI method which is used in the present work consists of an iterative
process in which a new and improved set of determinants is found
during each step. At each step, a new basis set of configurations is
picked by choosing the most significant configurations in the current
estimate for the wave function, and adding in any other configurations
that are strongly coupled through the Hamiltonian to these significant
configurations. An improved estimate for the wave function is then
obtained by diagonalizing the Hamiltonian in the basis of this newly
picked configuration basis, and the procedure is repeated until
convergence.  Only the expansion length is specified by the user, the
optimal structure of the wave function arising automatically during
the solution: significant determinants are generated, while
insignificant ones are ignored. ASCI is generally variational,
reproducing the best wave function with a given number of
determinants.\cite{Tubman2016}

Once the strong correlation in the CI wave function has been
sufficiently captured, dynamic correlation can be efficiently treated
with a perturbative approach,\cite{Evangelista2014, Tubman2016}
similarly to what is done e.g. in the traditional CASPT2 {[}CAS with
  second order perturbation theory{]} approach.\cite{Andersson1990,
  Andersson1992, Malmqvist2008}

\subsection{Coupled-cluster\label{sub:CoupledCluster}}

The idea in coupled-cluster (CC) theory\cite{Cizek1966} is to replace
the linear ansatz of \eqref{ciexp} with an exponential one
\begin{equation}
  \ket{\Psi} = e^{\hat{T}_1 + \hat{T}_2 + \dots} \ket{\Phi_0}, \label{eq:ccexp}
\end{equation}
where $\hat{T}_n$ are again $n$-fold excitation operators and
intermediate normalization is used. The difference between
\eqref{ciexp,ccexp} is that the CC expansion retains size consistency
even when truncated to e.g. single and double excitations from the
reference. This is easy to understand by Taylor expanding the
exponential in \eqref{ccexp}, which yields products of the $\hat{T}_n$
exitation operators, coupling up to arbitrarily high rank. Thanks to
this, while CC with single and double excitations (CCSD) is exact for
a single helium atom alike CISD, CCSD is also exact for an
\emph{arbitrary number} of non-interacting helium atoms, in contrast
to CISD. In addition to the \emph{connected} double excitation in
$\hat{T}_2$, CCSD also includes a \emph{disconnected} double
excitation arising from $\frac 1 2 \hat{T}_1^2$. The disconnected
excitations appear at all ranks, and increase in complexity with
increasing rank.

Thus, CC is a more compact way of parametrizing the wave function than
CI, as disconnected excitations appear from the Taylor expansion of
\eqref{ccexp}. The more sophisticated ansatz allows CC to converge faster
in the excitation amplitudes than CI does.\cite{Olsen1996,
  Bartlett2007} Unfortunately, the size consistency and faster
convergence come at the cost of a loss of variationality. Namely, as
even \emph{truncated versions} of variational CC scale exponentially,
the usual approach is to solve the CC energy and amplitudes by
projection\cite{Crawford2000, Bartlett2007}
\begin{eqnarray}
  E = & \braket{\Phi_0 | \hat{H} e^{\hat{T}} | \Phi_0}, \label{eq:CC-E} \\
  0 = & \braket{\Phi_i^a | e^{-\hat{T}} \hat{H} e^{\hat{T}} | \Phi_0}, \label{eq:CC-S} \\
  0 = & \braket{\Phi_{ij}^{ab} | e^{-\hat{T}} \hat{H} e^{\hat{T}} | \Phi_0}, \label{eq:CC-D} \\
  \vdots \nonumber
\end{eqnarray}
where \eqref{CC-S, CC-D} determine the single and double excitation
amplitudes, respectively, and $\Phi_i^a$ (and $\Phi_{ij}^{ab}$) are
singly (and doubly) excited determinants, with electrons $i$ (and $j$)
being promoted to unoccupied orbitals $a$ (and $b$). The projective CC
approaches are not guaranteed to yield an upper bound for the energy,
and indeed, projective CC truncated at low orders often underestimates
the energy in systems with significant strong correlation.

Variational behavior can still be recovered by increasing the maximum
excitation level, but this also leads to a significant increase in the
computational cost. But, despite the fast convergence of CC in
excitation rank, high-rank excitations may still be necessary even in
CC to properly describe systems with significant strong correlation
effects. Thus, while CC is very well adapted to compactly describing
dynamic (weak) correlations due to its cluster type product structure,
CC is not as well adapted to describing strong correlation, which has
motivated recently motivated alternative approaches such as CC valence
bond theory\cite{Small2009, *Small2011, *Small2012, *Small2017} and
projected Hartree--Fock theory\cite{Jimenez-Hoyos2012, *Qiu2016,
  *Qiu2017}. This problem is further complicated by the fact that
unfortunately, one cannot go very far in the CC expansion in practice:
to our knowledge, the highest-order available efficient dense-tensor
CC implementation is CC with single through quadruple excitations
(CCSDTQ) in the \textsc{NWChem} and \textsc{Aces II
}programs.\cite{ACESII, Valiev2010} A sparse-tensor implementation of
CC with single through hextuple excitations (CCSDTQ56) has also been
reported.\cite{Parkhill2010a}

\subsection{Motivation for cluster decomposition\label{sub:ClusterDecomposition}}

The feasibility of CC calculations is limited to somewhat low rank due
to the complexity of the solution of the CC amplitude equations.
Implementations of high-order CC theory are typically based on
conventional FCI programs,\cite{Olsen2000, Kallay2000, Kallay2001}
which operate on determinant strings,\cite{Knowles1984} making the CC
calculation as costly as a FCI calculation. In contrast to recently
developed adaptive FCI methods, the feasibility of an adaptive FCC
approach is at present unclear. A stochastic CC approach has been
suggested in the literature,\cite{Thom2010, Spencer2016} but it is
still reliant on string-based methods.

In CI, the amplitudes are easy to solve, involving simply the solution
of an eigenvalue equation. As discussed above, multiple adaptive FCI
approaches have been recently suggested, and shown to be quite
powerful for the description of challenging multiconfigurational
systems. However, due to the contamination in the CI amplitudes by
disconnected diagrams, it would be interesting to see if one could
rewrite the selected CI wave functions in CC form and thereby extract
better information on the underlying correlations in the system.

As an example, orbital optimization in SCF theory corresponds to
$e^{\hat{T}_{1}}$ according to Thouless'
theorem\cite{Thouless1960}. Thus, while non-optimal reference orbitals
only appear in $\hat{T}_{1}$, in the CI expansion they will show up
not only in $\hat{C}_{1}$, but in all ranks of excitations
$\hat{C}_{n}$, which complicates the analysis of the excitation
amplitudes.  Thus, a poor choice of orbitals may result in all
$\{\hat{C}_{n}\}_{n=1}^{2N}$ being significant for a system of $N$
non-interacting helium atoms, while only $\hat{T}_{1}$ and
$\hat{T}_{2}$ are necessary for obtaining the exact solution of this
system.

Being able to break down the FCI wave function into a cluster
decomposition that unambiguously shows the importance of irreducible
higher order excitations would clearly be useful first to understand
the physics and chemistry of systems with complicated electronic
structure, and second to point the way to building better scaling
electronic structure models than the (quasi)exponentially scaling
FCI. Cluster amplitudes extracted from FCI might be used for treating
dynamical correlation within a CC-type approach,\cite{Kinoshita2005}
or help in developing novel types of truncations of CC theory for the
treatment of strong correlations.\cite{Parkhill2009, Parkhill2010,
  Parkhill2010b, Parkhill2011, Lehtola2016b, Lehtola2017}

While a special case of the cluster decomposition has been
published,\cite{Paldus1982} we are not aware of any general
implementation thereof. In the present manuscript we present a general
implementation of the cluster decomposition, applicable to any
FCI-type wave function in a Slater determinant basis -- including CAS,
GAS as well as selected CI approaches -- and apply it to both FCI wave
and ASCI wave functions.

The organization of the manuscript is the following. Next, in the
Theory section, we briefly review the mathematics of the cluster
decomposition.  Then, in the Implementation section, the approach used
to implement the procedure is described. Details of the calculations
are described in the Computational Details section. Various
applications of the procedure are shown in the Results section. The
article terminates with a short Summary and Discussion section.

\section{Theory\label{sec:Theory}}

As the exact wave function is the same regardless of the
parametrization used, from \eqref{ciexp, ccexp} we get the connection
between the CI and CC amplitudes as
\begin{align}
\exp\hat{T}= & 1+\hat{C}.\label{eq:ccci}
\end{align}
\noindent Taylor expanding the exponential operator in \eqref{ccci}
\begin{align}
\hat{T}+\frac{1}{2}\hat{T}^{2}+\frac{1}{3!}\hat{T}^{3}+\dots= & \hat{C}\label{eq:match-coeffs}
\end{align}
\noindent the excitation can be matched rank by rank to yield
\begin{align}
\hat{T}_{1}= & \hat{C}_{1},\label{eq:singles-orig}\\
\frac{1}{2}\hat{T}_{1}^{2}+\hat{T}_{2}= & \hat{C}_{2},\label{eq:doubles-orig}\\
\frac{1}{3!}\hat{T}_{1}^{3}+\frac{1}{2}\left(\hat{T}_{1}\hat{T}_{2}+\hat{T}_{2}\hat{T}_{1}\right)+\hat{T}_{3}= & \hat{C}_{3},\label{eq:triples-orig}\\
\vdots\nonumber
\end{align}
as has been discussed by Monkhorst.\cite{Monkhorst1977} Next, the
individual excitation operators can be written out explicitly in
second quantized form as

\begin{align}
\hat{O}_{n}= & \frac{1}{\left(n!\right)^{2}}\sum_{i_{1}a_{1}\dots i_{n}a_{n}}o_{i_{1}\dots i_{n}}^{a_{1}\dots a_{n}}a_{a_{1}}^{\dagger}\dots a_{a_{n}}^{\dagger}a_{i_{n}}\dots a_{i_{1}},\label{eq:op-red}
\end{align}
and the equations for the individual amplitudes be determined from the
expressions by projection onto the relevant determinant,
e.g.\begin{widetext}
\begin{align}
c_{ij}^{ab}= & \braket{\Phi_{ij}^{ab}|\hat{C}_{2}|\Phi_{0}}=\braket{\Phi_{ij}^{ab}|\hat{T}_{2}+\frac{1}{2}\hat{T}_{1}^{2}|\Phi_{0}}=\braket{a_{a}^{\dagger}a_{b}^{\dagger}a_{j}a_{i}\Phi_{0}|\frac{1}{4}t_{kl}^{cd}a_{c}^{\dagger}a_{d}^{\dagger}a_{l}a_{k}+\frac{1}{2}\left(t_{k}^{c}a_{c}^{\dagger}a_{k}\right)\left(t_{l}^{d}a_{d}^{\dagger}a_{l}\right)|\Phi_{0}}\nonumber \\
= & \braket{\Phi_{0}|a_{i}^{\dagger}a_{j}^{\dagger}a_{b}a_{a}\left[\frac{1}{4}t_{kl}^{cd}a_{c}^{\dagger}a_{d}^{\dagger}a_{l}a_{k}+\frac{1}{2}\left(t_{k}^{c}a_{c}^{\dagger}a_{k}\right)\left(t_{l}^{d}a_{d}^{\dagger}a_{l}\right)\right]|\Phi_{0}}=t_{ij}^{ab}+t_{i}^{a}t_{j}^{b}-t_{i}^{b}t_{j}^{a},\label{eq:C2}
\end{align}
\end{widetext}where the operator strings in the equations can easily
be evaluated with e.g. Wick contractions.\cite{Wick1950, Crawford2000}
The resulting equations are easily solved to obtain recursive
equations for the $t$ amplitudes: $T_n$ is obtained from $C_n$ by
removing products of $T_l$ with $l=1,\dots,n-1$.

Interestingly, the factors $1/n!$ in the Taylor expansion are
cancelled out by the number of tensor permutations within a given
term. For example, in the equation for $t_{ijk}^{abc}$, the Wick
contractions of $\hat{T}_{1}\hat{T}_{2}$ yield both
$t_{i}^{a}t_{jk}^{bc}$ and $t_{jk}^{bc}t_{i}^{a}$, exactly canceling
out the prefactor $1/2$ of the term. Thus, the only thing that is left
over are terms in which indices are swapped between different
amplitude tensors, that is, the result of the Wick contraction can be
written as an antisymmetrizer.

Finally, the elemental forms for the recursion relations are obtained
as

\begin{align}
t_{i}^{a}= & c_{i}^{a}\label{eq:S}\\
t_{ij}^{ab}= & c_{ij}^{ab}-t_{i}^{a}t_{j}^{b}+t_{i}^{b}t_{j}^{a}\label{eq:D}\\
t_{ijk}^{abc}= & c_{ijk}^{abc}-t_{i}^{a}t_{jk}^{bc}+t_{i}^{b}t_{jk}^{ac}-t_{i}^{c}t_{jk}^{ab}+t_{j}^{a}t_{ik}^{bc}-t_{j}^{b}t_{ik}^{ac}\nonumber \\
+ & t_{j}^{c}t_{ik}^{ab}-t_{k}^{a}t_{ij}^{bc}+t_{k}^{b}t_{ij}^{ac}-t_{k}^{c}t_{ij}^{ab}-t_{i}^{a}t_{j}^{b}t_{k}^{c}\nonumber \\
+ & t_{i}^{a}t_{j}^{c}t_{k}^{b}+t_{i}^{b}t_{j}^{a}t_{k}^{c}-t_{i}^{b}t_{j}^{c}t_{k}^{a}-t_{i}^{c}t_{j}^{a}t_{k}^{b}+t_{i}^{c}t_{j}^{b}t_{k}^{a}\label{eq:T}
\end{align}

\begin{widetext} for the singles, doubles, and triples. For the
quadruples we get

\begin{align}
t_{ijkl}^{abcd} & =c_{ijkl}^{abcd}-t_{i}^{a}t_{jkl}^{bcd}+t_{i}^{b}t_{jkl}^{acd}-t_{i}^{c}t_{jkl}^{abd}+t_{i}^{d}t_{jkl}^{abc}+t_{j}^{a}t_{ikl}^{bcd}-t_{j}^{b}t_{ikl}^{acd}+t_{j}^{c}t_{ikl}^{abd}-t_{j}^{d}t_{ikl}^{abc}-t_{k}^{a}t_{ijl}^{bcd}+t_{k}^{b}t_{ijl}^{acd}-t_{k}^{c}t_{ijl}^{abd}\nonumber \\
 & +t_{k}^{d}t_{ijl}^{abc}+t_{l}^{a}t_{ijk}^{bcd}-t_{l}^{b}t_{ijk}^{acd}+t_{l}^{c}t_{ijk}^{abd}-t_{l}^{d}t_{ijk}^{abc}-t_{ij}^{ab}t_{kl}^{cd}+t_{ij}^{ac}t_{kl}^{bd}-t_{ij}^{ad}t_{kl}^{bc}-t_{ij}^{bc}t_{kl}^{ad}+t_{ij}^{bd}t_{kl}^{ac}-t_{ij}^{cd}t_{kl}^{ab}+t_{ik}^{ab}t_{jl}^{cd}\nonumber \\
 & -t_{ik}^{ac}t_{jl}^{bd}+t_{ik}^{ad}t_{jl}^{bc}+t_{ik}^{bc}t_{jl}^{ad}-t_{ik}^{bd}t_{jl}^{ac}+t_{ik}^{cd}t_{jl}^{ab}-t_{il}^{ab}t_{jk}^{cd}+t_{il}^{ac}t_{jk}^{bd}-t_{il}^{ad}t_{jk}^{bc}-t_{il}^{bc}t_{jk}^{ad}+t_{il}^{bd}t_{jk}^{ac}-t_{il}^{cd}t_{jk}^{ab}-t_{i}^{a}t_{j}^{b}t_{kl}^{cd}\nonumber \\
 & +t_{i}^{a}t_{j}^{c}t_{kl}^{bd}-t_{i}^{a}t_{j}^{d}t_{kl}^{bc}+t_{i}^{a}t_{k}^{b}t_{jl}^{cd}-t_{i}^{a}t_{k}^{c}t_{jl}^{bd}+t_{i}^{a}t_{k}^{d}t_{jl}^{bc}-t_{i}^{a}t_{l}^{b}t_{jk}^{cd}+t_{i}^{a}t_{l}^{c}t_{jk}^{bd}-t_{i}^{a}t_{l}^{d}t_{jk}^{bc}+t_{i}^{b}t_{j}^{a}t_{kl}^{cd}-t_{i}^{b}t_{j}^{c}t_{kl}^{ad}\nonumber \\
 & +t_{i}^{b}t_{j}^{d}t_{kl}^{ac}-t_{i}^{b}t_{k}^{a}t_{jl}^{cd}+t_{i}^{b}t_{k}^{c}t_{jl}^{ad}-t_{i}^{b}t_{k}^{d}t_{jl}^{ac}+t_{i}^{b}t_{l}^{a}t_{jk}^{cd}-t_{i}^{b}t_{l}^{c}t_{jk}^{ad}+t_{i}^{b}t_{l}^{d}t_{jk}^{ac}-t_{i}^{c}t_{j}^{a}t_{kl}^{bd}+t_{i}^{c}t_{j}^{b}t_{kl}^{ad}-t_{i}^{c}t_{j}^{d}t_{kl}^{ab}\nonumber \\
 & +t_{i}^{c}t_{k}^{a}t_{jl}^{bd}-t_{i}^{c}t_{k}^{b}t_{jl}^{ad}+t_{i}^{c}t_{k}^{d}t_{jl}^{ab}-t_{i}^{c}t_{l}^{a}t_{jk}^{bd}+t_{i}^{c}t_{l}^{b}t_{jk}^{ad}-t_{i}^{c}t_{l}^{d}t_{jk}^{ab}+t_{i}^{d}t_{j}^{a}t_{kl}^{bc}-t_{i}^{d}t_{j}^{b}t_{kl}^{ac}+t_{i}^{d}t_{j}^{c}t_{kl}^{ab}-t_{i}^{d}t_{k}^{a}t_{jl}^{bc}\nonumber \\
 & +t_{i}^{d}t_{k}^{b}t_{jl}^{ac}-t_{i}^{d}t_{k}^{c}t_{jl}^{ab}+t_{i}^{d}t_{l}^{a}t_{jk}^{bc}-t_{i}^{d}t_{l}^{b}t_{jk}^{ac}+t_{i}^{d}t_{l}^{c}t_{jk}^{ab}-t_{j}^{a}t_{k}^{b}t_{il}^{cd}+t_{j}^{a}t_{k}^{c}t_{il}^{bd}-t_{j}^{a}t_{k}^{d}t_{il}^{bc}+t_{j}^{a}t_{l}^{b}t_{ik}^{cd}-t_{j}^{a}t_{l}^{c}t_{ik}^{bd}\nonumber \\
 & +t_{j}^{a}t_{l}^{d}t_{ik}^{bc}+t_{j}^{b}t_{k}^{a}t_{il}^{cd}-t_{j}^{b}t_{k}^{c}t_{il}^{ad}+t_{j}^{b}t_{k}^{d}t_{il}^{ac}-t_{j}^{b}t_{l}^{a}t_{ik}^{cd}+t_{j}^{b}t_{l}^{c}t_{ik}^{ad}-t_{j}^{b}t_{l}^{d}t_{ik}^{ac}-t_{j}^{c}t_{k}^{a}t_{il}^{bd}+t_{j}^{c}t_{k}^{b}t_{il}^{ad}-t_{j}^{c}t_{k}^{d}t_{il}^{ab}\nonumber \\
 & +t_{j}^{c}t_{l}^{a}t_{ik}^{bd}-t_{j}^{c}t_{l}^{b}t_{ik}^{ad}+t_{j}^{c}t_{l}^{d}t_{ik}^{ab}+t_{j}^{d}t_{k}^{a}t_{il}^{bc}-t_{j}^{d}t_{k}^{b}t_{il}^{ac}+t_{j}^{d}t_{k}^{c}t_{il}^{ab}-t_{j}^{d}t_{l}^{a}t_{ik}^{bc}+t_{j}^{d}t_{l}^{b}t_{ik}^{ac}-t_{j}^{d}t_{l}^{c}t_{ik}^{ab}-t_{k}^{a}t_{l}^{b}t_{ij}^{cd}\nonumber \\
 & +t_{k}^{a}t_{l}^{c}t_{ij}^{bd}-t_{k}^{a}t_{l}^{d}t_{ij}^{bc}+t_{k}^{b}t_{l}^{a}t_{ij}^{cd}-t_{k}^{b}t_{l}^{c}t_{ij}^{ad}+t_{k}^{b}t_{l}^{d}t_{ij}^{ac}-t_{k}^{c}t_{l}^{a}t_{ij}^{bd}+t_{k}^{c}t_{l}^{b}t_{ij}^{ad}-t_{k}^{c}t_{l}^{d}t_{ij}^{ab}+t_{k}^{d}t_{l}^{a}t_{ij}^{bc}-t_{k}^{d}t_{l}^{b}t_{ij}^{ac}\nonumber \\
 & +t_{k}^{d}t_{l}^{c}t_{ij}^{ab}-t_{i}^{a}t_{j}^{b}t_{k}^{c}t_{l}^{d}+t_{i}^{a}t_{j}^{b}t_{k}^{d}t_{l}^{c}+t_{i}^{a}t_{j}^{c}t_{k}^{b}t_{l}^{d}-t_{i}^{a}t_{j}^{c}t_{k}^{d}t_{l}^{b}-t_{i}^{a}t_{j}^{d}t_{k}^{b}t_{l}^{c}+t_{i}^{a}t_{j}^{d}t_{k}^{c}t_{l}^{b}+t_{i}^{b}t_{j}^{a}t_{k}^{c}t_{l}^{d}-t_{i}^{b}t_{j}^{a}t_{k}^{d}t_{l}^{c}\nonumber \\
 & -t_{i}^{b}t_{j}^{c}t_{k}^{a}t_{l}^{d}+t_{i}^{b}t_{j}^{c}t_{k}^{d}t_{l}^{a}+t_{i}^{b}t_{j}^{d}t_{k}^{a}t_{l}^{c}-t_{i}^{b}t_{j}^{d}t_{k}^{c}t_{l}^{a}-t_{i}^{c}t_{j}^{a}t_{k}^{b}t_{l}^{d}+t_{i}^{c}t_{j}^{a}t_{k}^{d}t_{l}^{b}+t_{i}^{c}t_{j}^{b}t_{k}^{a}t_{l}^{d}-t_{i}^{c}t_{j}^{b}t_{k}^{d}t_{l}^{a}-t_{i}^{c}t_{j}^{d}t_{k}^{a}t_{l}^{b}\nonumber \\
 & +t_{i}^{c}t_{j}^{d}t_{k}^{b}t_{l}^{a}+t_{i}^{d}t_{j}^{a}t_{k}^{b}t_{l}^{c}-t_{i}^{d}t_{j}^{a}t_{k}^{c}t_{l}^{b}-t_{i}^{d}t_{j}^{b}t_{k}^{a}t_{l}^{c}+t_{i}^{d}t_{j}^{b}t_{k}^{c}t_{l}^{a}+t_{i}^{d}t_{j}^{c}t_{k}^{a}t_{l}^{b}-t_{i}^{d}t_{j}^{c}t_{k}^{b}t_{l}^{a}\label{eq:Q}.
\end{align}

\end{widetext}

\section{Implementation\label{sec:Implementation}}

We have written a C++ program which generates the recursion routines
by Wick contractions of antisymmetrized products. As is already
obvious from \eqrangeref{S}{Q}, the total number of contractions
increases rapidly with tensor rank, as is further illustrated in
\tabref{Number-of-terms}. Due to the rapidly increasing computational
requirements, we have only generated the recursion relations up to
octuple excitations, for which the instructions in text format take
roughly 500 megabytes.  The number of terms in the recursion relations
can be analysed by studying the ratio of consecutive ranks
\begin{align}
f(n) & =N(n)/N(n-1).\label{eq:fn}
\end{align}
For an exponentially scaling $N(n)\propto a^{n}$ the ratio is
constant, $f(n)=a$. However, examination of the actual scaling of the
number of terms with increasing rank (\tabref{Number-of-terms})
reveals that the decomposition scales \emph{superexponentially} with
respect to its rank. The reason for this, of course, is the
exponential scaling wall of the CI problem. However, the decomposition
scales only \emph{linearly} with respect to the length of the CI
expansion, even though the time to decompose a given determinant
(i.e. the prefactor for the linear scaling) depends on its rank.

\begin{table}
\begin{centering}
\begin{tabular}{lll}
$n$ & $N(n)$ & $f(n)$\tabularnewline
\midrule
\midrule
2 & 2 & \tabularnewline
3 & 15 & 7.50\tabularnewline
4 & 130 & 8.67\tabularnewline
5 & 1495 & 11.5\tabularnewline
6 & 22481 & 15.0\tabularnewline
7 & 426832 & 19.0\tabularnewline
8 & 9934562 & 23.3\tabularnewline
\end{tabular}
\par\end{centering}
\caption{Number of disconnected terms $N$ contained in $C_{n}$ and the resulting
ratio $f(n)$.\label{tab:Number-of-terms}}
\end{table}

While low-rank implementations of the cluster decomposition could be
generated on-the-fly and/or implemented in computer source code,
already for hextuples the latter becomes infeasible due to the large
size of the generated source. For higher ranks, also the former
approach -- the generation of the recursion relations -- becomes
costly. Because of this, we have chosen to generate and store the
recursion relations on disk in text format. The relations can then be
read in by a separate program that runs the decomposition in a
spin-orbital basis. The \textsc{ClusterDec} suite,\cite{clusterdec}
composed of the generator and the runtime programs, is freely
available on the internet.

A mathematically correct implementation of the cluster decomposition
of a given CI wave function would require the full evaluation of the
elements of $T_n$. However, this would require the calculation and
storage of $o^n v^n$ elements for a rank-$n$ expansion, quickly
exhausting any computational resources available to the user. But, the
very same scaling problem exists also for the original FCI problem,
where it has been solved in the adaptive and stochastic approaches by
using a sparse representation for $C_n$ instead of storing all its
$o^n v^n$ elements. As our main goal is to use the decomposition with
sparse FCI vectors, the implementation in \textsc{ClusterDec} uses the
same sparse representation for $\hat{T}_{n}$ as is used in the input
wave function $\hat{C}_{n}$. In other words, we assume that
$\hat{T}_{l}$ is fully spanned by the same subspace as $\hat{C_{l}}$
that is found variationally by the adaptive FCI routines.

This procedure is mathematically non-rigorous, as
$c_{ijk\dots}^{abc\dots} = t_{ijk\dots}^{abc\dots} + \text{products of
  lower-order t's} = \text{connected} + \text{disconnected
  diagrams}$. Thus, a negligible $c_{ijk\dots}^{abc\dots}$ (that is
omitted from the truncated CI expansion) does not necessarily imply
that the matching $t_{ijk\dots}^{abc\dots}$ is also negligible, as it
is possible that the CI expansion coefficient vanishes due to
destructive interference of the connected and disconnected diagrams.

However, this approximation is not only necessary for a practical
implementation, but also makes physical sense. Consider, for example,
the system of $n$ non-interacting helium atoms discussed in the
Introduction. Again, the FCC wave function will be fully spanned
by $T_1$ and $T_2$, with $T_n=0$ for $n\geq 3$. In contrast, the CI
wave function may contain non-zero $C_n$ for $n \in [1,2n]$. Thus, the
CI wave function will especially contain considerable contributions
from $C_1 = T_1$ and $C_2 = T_2 + \frac 1 2 T_1^2$, leading to their
inclusion in the adaptive wave function expansion, which also will
result in the accurate description of $T_1$ and $T_2$ in the truncated
basis. More generally, owing to the much slower convergence rate of
the CI expansion compared to that in the CC expansion,\cite{Olsen1996,
  Bartlett2007} the parameter space of the CI wave function is much
larger than that of the CC expansion, as most of the excitations in a
CI wave function for a system of non-interacting subsystems are pure
disconnected diagrams, and thus a converged CI expansion should
guarantee that the same basis can be used for performing a cluster
decomposition.  At the other extreme, in which the disconnected
diagrams play only a small role, the CI and CC expansions are
numerically similar, and the truncated CI expansion is again an
excellent basis for truncating the CC expansion.

Real calculations are somewhere in-between the two
extremes. Disconnected diagrams become dominant in extended systems,
as electron correlation is well known to be a local phenomenon. While
we have imposed a mathematically non-rigorous truncation of the CC
expansion, the exact decomposition can in principle always be
recovered in the limit of no truncation of the original CI wave
function, as this will also mean that the full determinant basis is
used for the cluster decomposition. In any case, even if the full
(i.e. non-sparse) cluster decomposition of a truncated CI wave
function were performed, it would still be necessary to check the
convergence of the decomposition with respect to the truncation of the
original CI wave function.

Thus, the rationale for the chosen tensor basis can be summarized in
two cases. If on the one hand a true FCI wave function is used, there
is no truncation in the cluster decomposition, because all elements of
the excitation tensors are spanned. On the other hand, if a truncated
CI wave function is used, then it might be the case that the CI wave
function itself is not sufficicently converged, and the convergence
with respect to the expansion length must be checked. As the CI wave
function is pushed towards convergence, the basis for the cluster
decomposition is also being improved towards convergence. For the
physical reasons detailed in the preceding paragraphs, the CI wave
function is less sparse than the CC wave function, and thus the error
from the incomplete decomposition basis for the CC coefficients
derived from an incomplete CI expansion is much smaller than the error
in the original CI wave function itself. A convergence check for the
CI expansion will be always necessary, as convergence of the energy
does not imply sufficient convergence of the wave function.

The importance of connected and disconnected contributions in the CI
amplitudes can be examined by studying the ratio $r_l = \left\Vert
T_{l}\right\Vert /\left\Vert C_{l}\right\Vert $, where the usual
$\mathcal{L}^{2}$ norm is used
\begin{align*}
\left\Vert O_{n}\right\Vert = & \sqrt{\sum_{i_{1}\dots i_{n},a_{1}\dots a_{n}}\left(o_{i_{1}\dots i_{n}}^{a_{1}\dots a_{n}}\right)^{2}}.
\end{align*}
In the case where the disconnected excitations are dominant, this
ratio is smaller than one: $r_l \ll 1$. This is easily the case when
the connected diagrams are small in amplitude; that is, when the
correlations are weak, or, of a dynamical nature. Because the CC
hierarchy recovers correlation effects much faster than the CI
expansion -- especially in extended systems -- the ratio should thus
decrease rapidly with increasing $l$. Cases where the connected
excitations are dominant, the ratio is close to one $r_l \approx
1$. Finally, cases in which there is destructive interference in the
CI coefficients are revealed by $r_l > 1$.

The excitation amplitudes are stored in {\sc ClusterDec} in bit string
form as
\begin{align}
\hat{O}_{n}= & \sum_{\substack{i_{1}<\dots<i_{n}\\
a_{1}<\dots<a_{n}
}
}o_{i_{1}\dots i_{n}}^{a_{1}\dots a_{n}}a_{a_{1}}^{\dagger}\dots a_{a_{n}}^{\dagger}a_{i_{n}}\dots a_{i_{1}},\label{eq:op-nonred}
\end{align}
similarly to what is done in FCI approaches.\cite{Knowles1984} To
extract elements of the CI amplitude tensor, such as $c_{i}^{a}$, we
note that it is necessary to project the CI wave function $\ket{\Psi}$
onto the wanted determinant $\ket{\Phi_i^a}$, as the Wick contraction
of the bit strings may contribute a change of sign. For instance, for a
$(2e,2o)$ problem with orbitals $1$ and $2$ occupied and $3$ and $4$
unoccupied in the reference determinant, we have
\begin{align*}
\braket{\Phi_{2}^{3}|a_{1}^{\dagger}a_{3}^{\dagger}0}= & \braket{a_{3}^{\dagger}a_{2}\Phi_{0}|a_{1}^{\dagger}a_{3}^{\dagger}0}\\
= & \braket{a_{3}^{\dagger}a_{2}a_{1}^{\dagger}a_{2}^{\dagger}0|a_{1}^{\dagger}a_{3}^{\dagger}0}\\
= & \braket{0|a_{2}a_{1}a_{2}^{\dagger}a_{3}a_{1}^{\dagger}a_{3}^{\dagger}0}=1
\end{align*}
while
\begin{align*}
\braket{\Phi_{1}^{4}|a_{2}^{\dagger}a_{4}^{\dagger}0}= & \braket{a_{4}^{\dagger}a_{1}\Phi_{0}|a_{2}^{\dagger}a_{4}^{\dagger}0}\\
= & \braket{a_{4}^{\dagger}a_{1}a_{1}^{\dagger}a_{2}^{\dagger}0|a_{2}^{\dagger}a_{4}^{\dagger}0}\\
= & \braket{0|a_{2}a_{1}a_{1}^{\dagger}a_{4}a_{2}^{\dagger}a_{4}^{\dagger}0}=-1.
\end{align*}
The necessary sign for any determinant string is easily found using
automatical Wick contraction routines.

\section{Computational details\label{sec:Computational-details}}

Two types of FCI wave functions are considered. True FCI wave
functions, obtained via the diagonalization of the molecular
Hamiltonian operator in the full determinant space, are calculated
using \textsc{Psi4}.\cite{Sherrill1999, Turney2011}

In cases where the FCI function would be too large, selected CI wave
functions are computed using the ASCI approach,\cite{Tubman2016} which
tries to find the best possible compact approximation to the true FCI
wave function by treating only its most important configurations. The
ASCI method is an iterative process, which can be summarized as
follows:
\begin{enumerate}
\item[0.] Set the starting wave function which is defined with a set
  of coefficents $\{C_k\}$ and corresponding determinants
  $\{\ket{\Phi_k}\}$.
\item[1.] Estimate expansion coefficients for additional
  configurations as\cite{Tubman2016}
\begin{equation}
 A_{i} =\frac{\sum_{j \ne i}^\text{core} H_{ij} C_{j} } {H_{ii}-E}, \label{eqn:ranking}
\end{equation}
over all single and double substitutions $D^{SD}$ from the
configurations $\{\ket{\Phi_k}\}$ in the current wave function, where
$H_{ij} = \braket{\Phi_i|H|\Phi_j}$.
\item[2.] Use the largest values of $\{A_i\}$ from the singles
  and doubles space and $\{C_i\}$ from the core space to rank and
  select a new set of determinants, denoted as $\{\Phi^{1}\}$.  
  %Generally equation \ref{eqn:ranking} is calculated in an approximate manner with modern selected CI approaches.
\item[3.] Form the new matrix elements for the Hamiltonian in the
  basis of the determinants $\{\Phi^{1}\}$, diagonalize, and use the
  resulting wave function as input for the next iteration.
\end{enumerate}
At each iteration, an improved set of configurations is picked by
choosing the most significant configurations in the current estimate
for the wave function according to step 2, and adding in any other
configurations that are strongly coupled through the Hamiltonian to
these significant configurations. A new wave function is then obtained
by diagonalizing the Hamiltonian in the basis of these newly picked
configurations, and the procedure is repeated until convergence.

In an ASCI calculation performed using $N$ determinants in the
diagonalization, all but the last coefficients of the expansion are
typically found to be converged with respect to the true FCI wave
function.\cite{Tubman2016} Details on the algorithms and
implementation of ASCI are available elsewhere.\cite{Tubman2016,
  Holmes2016a}

For the simulations performed in this work, we generated very large
ASCI wave functions to help ensure that we were generating the top
determinants for use in the cluster decomposition.  In most of the
simulations over $10^{7}$ determinants were generated. For purposes of
the cluster decomposition, we converged the parameter for our search
space, (referred to as \emph{cdets} in ref.~\cite{Tubman2016}), such
that the configurations that appear in the resulting wave function did
not change anymore upon a further increase of
\emph{cdets}. Convergence was generally found to be achieved with a
$cdets$ value in the range of 50,000 to 100,000.

\section{Results\label{sec:Results}}

To study the suitability of the cluster decomposition to truncated
wave functions with the further approximation of the sparse tensor
basis described in \secref{Implementation}, the FCI and ASCI wave
functions are analyzed by running the decomposition on a further
truncation of the wave function. This is done by only including the
first $n$ determinants from the fixed input wave function, sorted in
descending absolute values of the expansion coefficients. To simplify
the analysis, the truncated wave functions were not renormalized. The
bulk of the norm is typically contained within the first few dozen
determinants, which are always included in the analysis, thus
renormalization should have little to no effect on the results of the
analysis.

\subsection{Double dissociation of water\label{sub:DoubleDissociation}}

\subsubsection{FCI/6-31G}

\begin{figure*}
\subfloat[$r_{\text{OH}}^{0}$, first 1k\label{fig:h2o-eq-631g-1k}]{\includegraphics[width=0.25\textwidth]{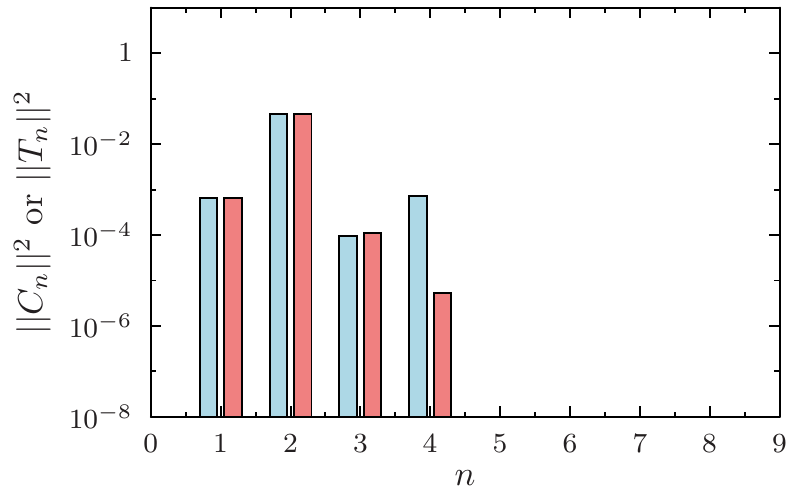}} \subfloat[$r_{\text{OH}}^{0}$, first 10k\label{fig:h2o-eq-631g-10k}]{\includegraphics[width=0.25\textwidth]{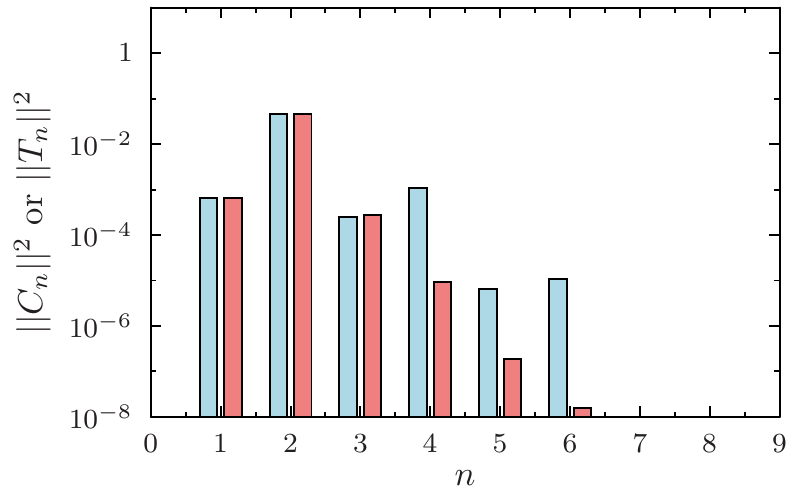}} \subfloat[$r_{\text{OH}}^{0}$, first 100k\label{fig:h2o-eq-631g-100k}]{\includegraphics[width=0.25\textwidth]{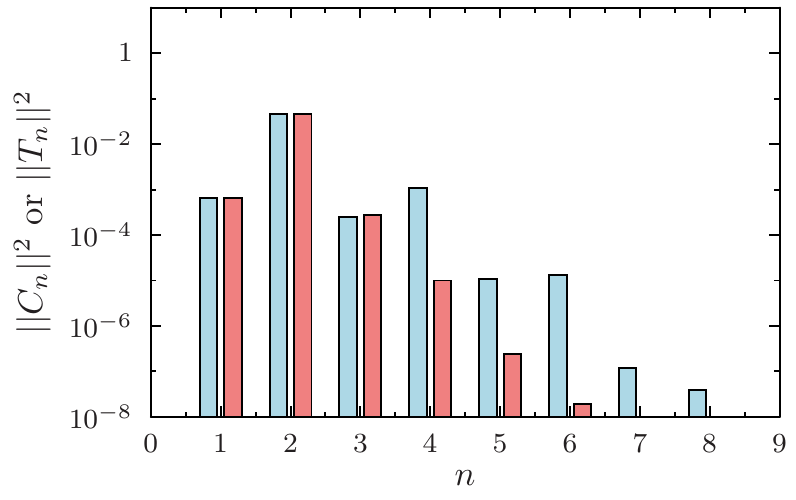}} \subfloat[$r_{\text{OH}}^{0}$, all\label{fig:h2o-eq-631g-all}]{\includegraphics[width=0.25\textwidth]{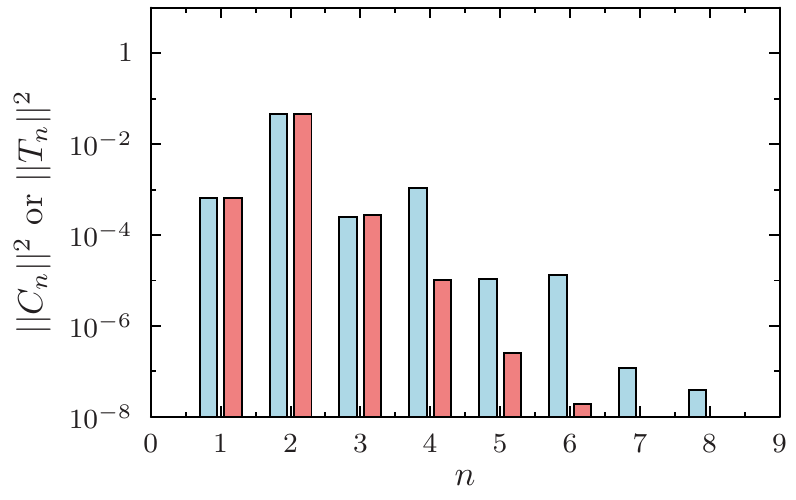}}

\subfloat[$2r_{\text{OH}}^{0}$, first 1k\label{fig:h2o-20-631g-1k}]{\includegraphics[width=0.25\textwidth]{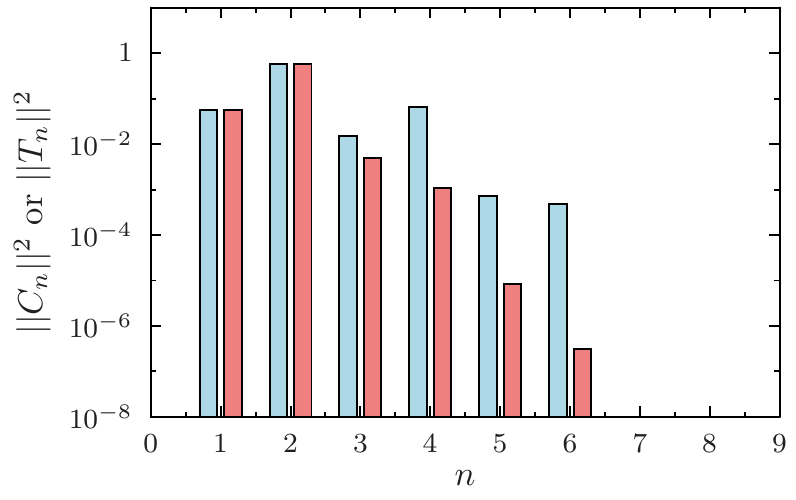}} \subfloat[$2r_{\text{OH}}^{0}$, first 10k\label{fig:h2o-20-631g-10k}]{\includegraphics[width=0.25\textwidth]{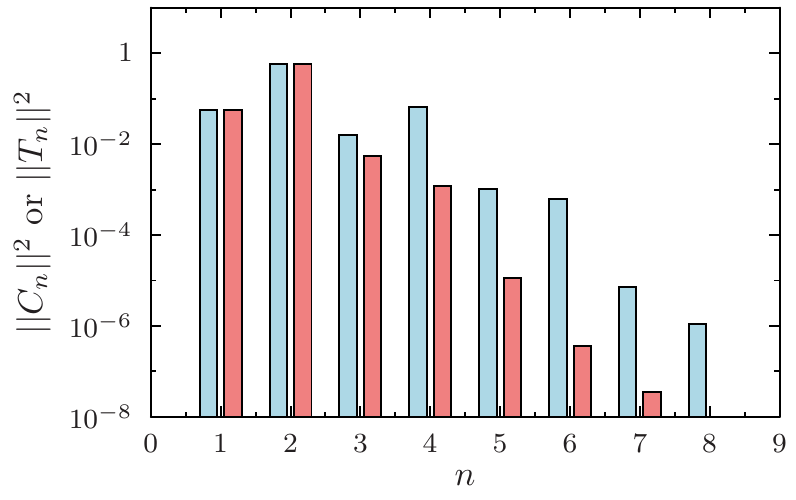}} \subfloat[$2r_{\text{OH}}^{0}$, first 100k\label{fig:h2o-20-631g-100k}]{\includegraphics[width=0.25\textwidth]{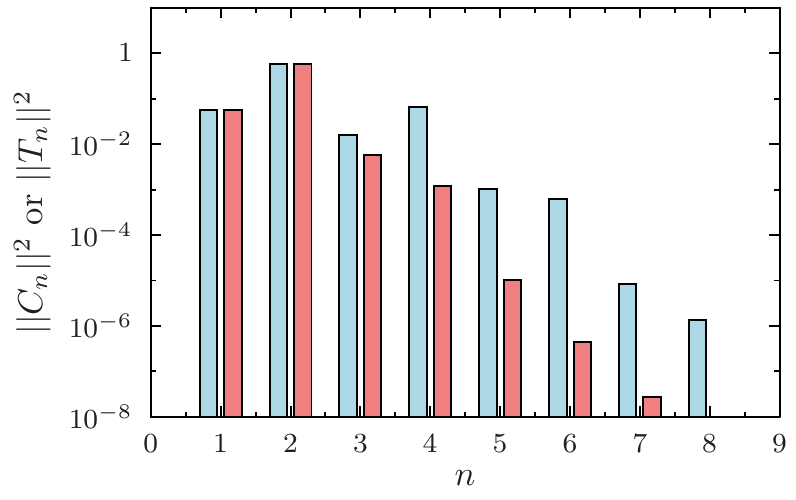}} \subfloat[$2r_{\text{OH}}^{0}$, all\label{fig:h2o-20-631g-all}]{\includegraphics[width=0.25\textwidth]{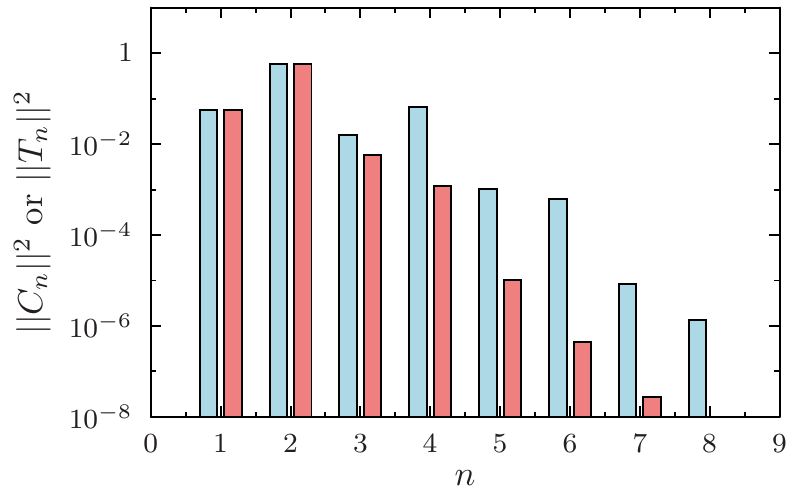}}

\subfloat[$3r_{\text{OH}}^{0}$, first 1k\label{fig:h2o-30-631g-1k}]{\includegraphics[width=0.25\textwidth]{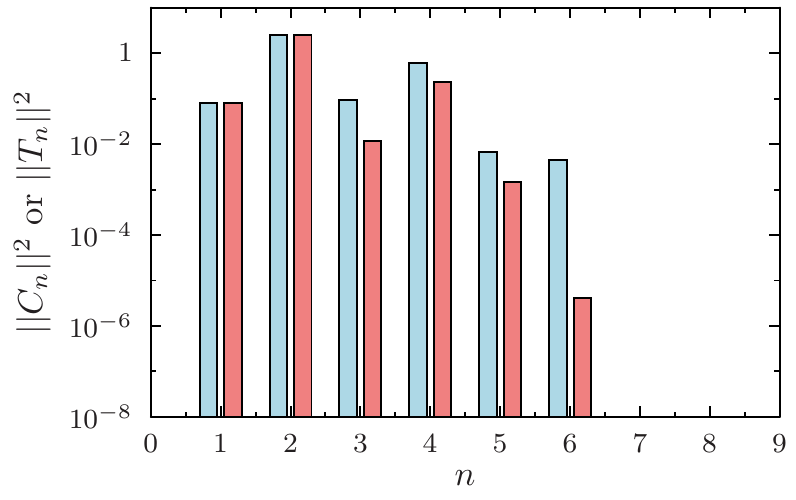}} \subfloat[$3r_{\text{OH}}^{0}$, first 10k\label{fig:h2o-30-631g-10k}]{\includegraphics[width=0.25\textwidth]{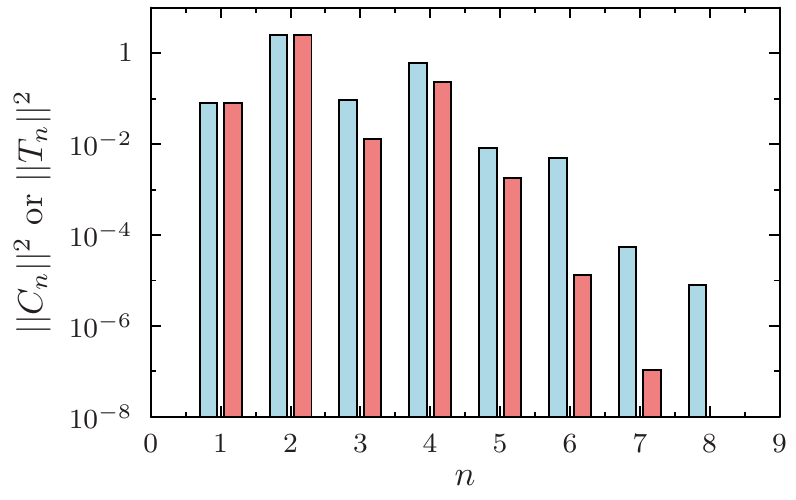}} \subfloat[$3r_{\text{OH}}^{0}$, first 100k\label{fig:h2o-30-631g-100k}]{\includegraphics[width=0.25\textwidth]{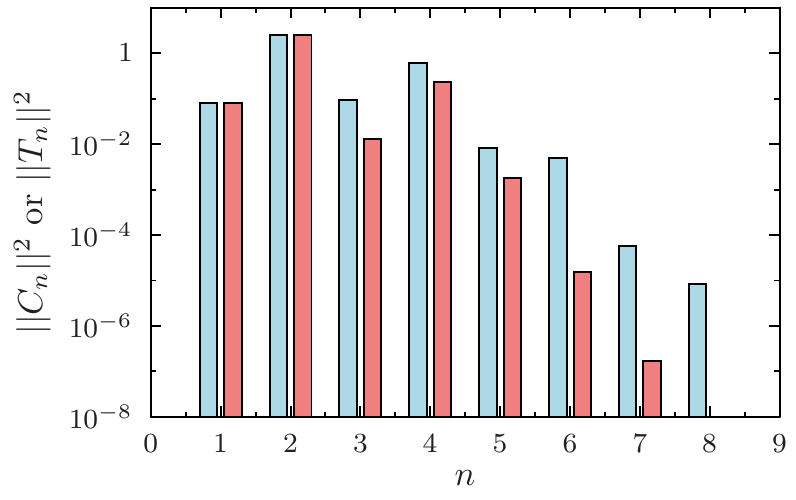}} \subfloat[$3r_{\text{OH}}^{0}$, all\label{fig:h2o-30-631g-all}]{\includegraphics[width=0.25\textwidth]{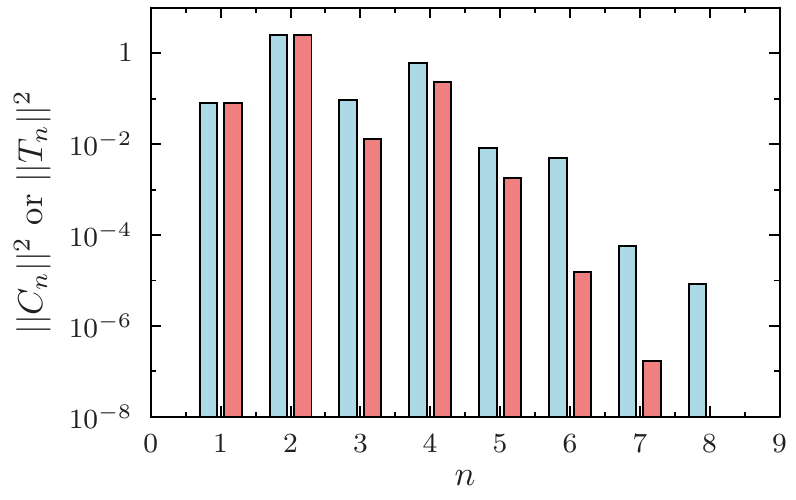}}

 \caption{CI (blue) and CC (red) coefficients for a FCI wave function
   of water in 6-31G basis at various OH bond lengths $r_{\text{OH}}$,
   plots with the first 1k, 10k, 100k, and all determinants,
   increasing from the left. Note logarithmic
   scale.\label{fig:Water-631g}}
\end{figure*}

The double dissociation of water has been studied by Olsen et al with
FCI and CC methods in the cc-pVDZ basis set.\cite{Olsen1996,
  Dunning1989} Here, we repeat the calculation in the 6-31G basis
set\cite{Hehre1972} where the FCI wave function contains 1656369
determinants. A restricted HF reference configuration is
used. Following Olsen et al, five different OH bond lengths are used,
with the equilibrium geometry taken as
$\angle(\text{HOH})=110.6^{\circ}$ and $r_{\text{OH}}^{0}=1.84345$
bohr.\cite{Olsen1996} Although the FCI problem is solved here in the
full determinant space, we study the suitability of wave function
truncation in \figref{Water-631g}. The weights of the HF reference and
FCI energies are given in \tabref{DDIS}.

The best way to start analyzing the results of \figref{Water-631g} is
through its rightmost column, which corresponds to the use of the full
FCI wave function, and thus contains no approximations at
all. \ce{H2O} at equilibrium is dominated by dynamic correlation
effects, which is clearly visible in the rapidly decaying cluster
expansion of \figref{h2o-eq-631g-all}.  For instance, inclusion of all
$T_{n}$ for $\left\Vert T_{n}\right\Vert \lesssim10^{-4}$ would
require CCSDT. Stretching the OH bond length towards double
dissociation monotonically significantly increases the importance of
the excitation operators, which is not explained simply by the change
of the amplitude of the HF configuration (see \tabref{DDIS}).  For the
most stretched geometry, the same criterion of $\left\Vert
T_{n}\right\Vert \lesssim10^{-4}$ would require CCSDTQ56. For
reference, results from calculations up to CCSDTQ5, performed with
\textsc{Q-Chem},\cite{Parkhill2010a, Shao2015} are shown in
\tabref{DDIS-CC}.

The minimal framework for describing the two single bonds in \ce{H2O}
is the (4e,4o) problem. This is seen at stretched geometries, as the
cluster decomposition shows a multi-peak structure with the largest
changes with respect to the decomposition at equilibrium being seen
for $T_{4}$. At $3r_{\text{OH}}^{0}$ (a 200\% stretched geometry),
quadruple excitations are the second most important amplitude, after
$T_{2}$. Note that both CCSD and CCSDT with restricted orbitals are
well known to undergo variational collapse for this system, CCSDTQ
being the first level of CC theory that predicts the qualitatively
correct dissociation curve. This insight into the system is evident
from the importance of $T_{4}$ in \figref{h2o-30-631g-all}, supporting
our expectation that the cluster analysis can be used to establish
requirements for obtaining converged results on challenging systems
with CC theory.

While one would expect CCSDTQ to be extremely accurate for the double
bond dissociation of the present example, the decomposition at
$3r_{\text{OH}}^{0}$ still has surprisingly large contributions from
$T_{n}$ for $n>4$. This indicates that the lone pairs on the oxygen
atom also participate in the strong correlation effects. This is also
clear from the truncated CC energies of \tabref{DDIS-CC}: the error at
CCSDTQ level is still several millihartrees.

Moving on to the truncated FCI wave functions, the decomposition using
only the first 100k determinants out of the 1.6M determinant wave
function is almost fully converged. Even the decomposition based only
on the first 10k determinants is qualitatively correct, with small differences
being seen for the higher excitations. These results show that a
converged cluster decomposition can be obtained even when a truncated
FCI wave function is used and when the decomposition is performed in a
sparse basis, so next we can look at real, approximate FCI wave
functions.

\begingroup
\squeezetable

\begin{table}
\begin{centering}
\begin{tabular}{cccc}
$r_{\text{OH}}/r_{\text{OH}}^{0}$ & $C_{0}$ & $E_{\text{HF}}/E_{H}$ & $E_{\text{FCI}}/E_{H}$\tabularnewline
\midrule
\midrule
1.0 & 0.977 & -75.984079 & -76.122302\tabularnewline
1.5 & 0.924 & -75.780587 & -75.980926\tabularnewline
2.0 & 0.765 & -75.573397 & -75.874634\tabularnewline
2.5 & 0.584 & -75.425644 & -75.843213\tabularnewline
3.0 & 0.483 & -75.327022 & -75.837391\tabularnewline
\end{tabular}
\par\end{centering}
\caption{Weights of the HF reference $C_{0}$ as well as HF and FCI energies
for the double dissociation of water in the 6-31G basis.\label{tab:DDIS}}
\end{table}

\begin{table}
\begin{centering}
\begin{tabular}{crrrr}
$r_{\text{OH}}/r_{\text{OH}}^{0}$ & $\Delta E_{\text{CCSD}}$ & $\Delta E_{\text{CCSDT}}$ & $\Delta E_{\text{CCSDTQ}}$ & $\Delta E_{\text{CCSDTQ5}}$\tabularnewline
\midrule
\midrule
1.0 & $1.6\times10^{-3}$ & $4.6\times10^{-4}$ & $1.2\times10^{-5}$ & $3.3\times10^{-6}$\tabularnewline
1.5 & $5.9\times10^{-3}$ & $1.2\times10^{-3}$ & $1.0\times10^{-4}$ & $1.5\times10^{-5}$\tabularnewline
2.0 & $1.0\times10^{-2}$ & $-2.5\times10^{-3}$ & $7.8\times10^{-5}$ & $2.1\times10^{-5}$\tabularnewline
2.5 & $-6.7\times10^{-3}$ & $-2.6\times10^{-2}$ & $-1.4\times10^{-3}$ & $-4.9\times10^{-5}$\tabularnewline
3.0 & $-2.1\times10^{-2}$ & $-4.2\times10^{-2}$ & $-2.9\times10^{-3}$ & NC\tabularnewline
\end{tabular}
\par\end{centering}
\caption{Discrepancy between FCI and CC energies for the double
  dissociation of water in the 6-31G basis. NC: no convergence
  achieved within the run time.\label{tab:DDIS-CC}}
\end{table}

\endgroup

\subsubsection{ASCI/cc-pVTZ}

\begin{figure*}
  \subfloat[$r_{\text{OH}}^{0}$, first 1k\label{fig:h2o-eq-cc-pvtz-1k}]{\includegraphics[width=0.25\textwidth]{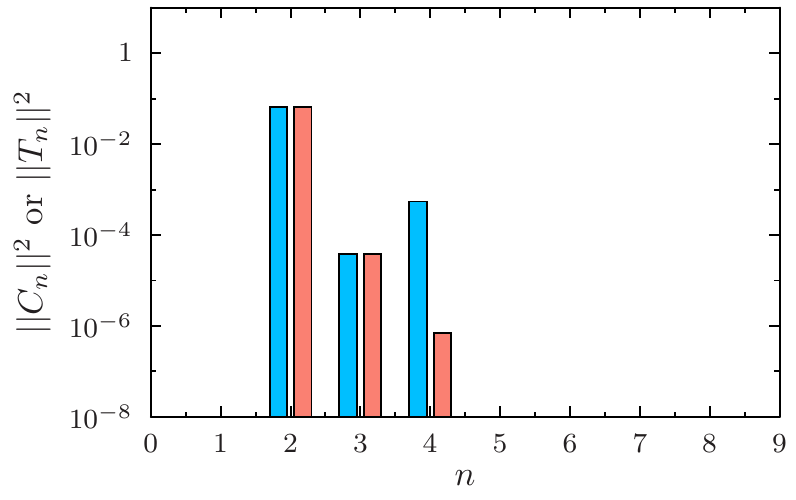}} \subfloat[$r_{\text{OH}}^{0}$, first 10k\label{fig:h2o-eq-cc-pvtz-10k}]{\includegraphics[width=0.25\textwidth]{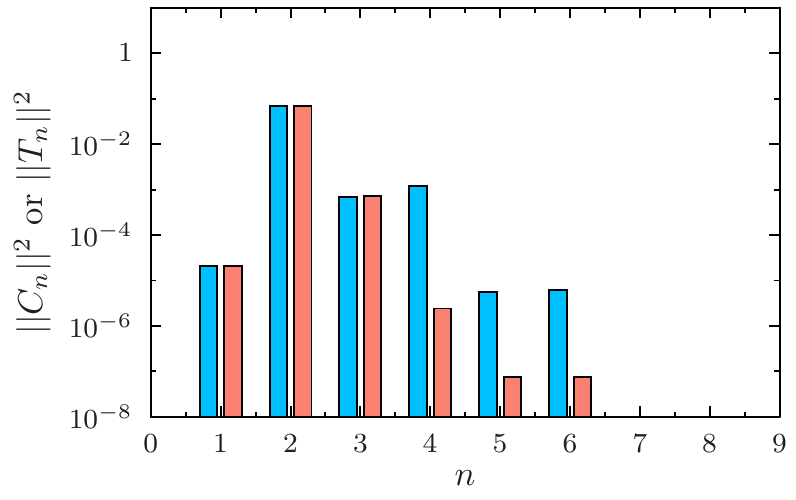}} \subfloat[$r_{\text{OH}}^{0}$, first 100k\label{fig:h2o-eq-cc-pvtz-100k}]{\includegraphics[width=0.25\textwidth]{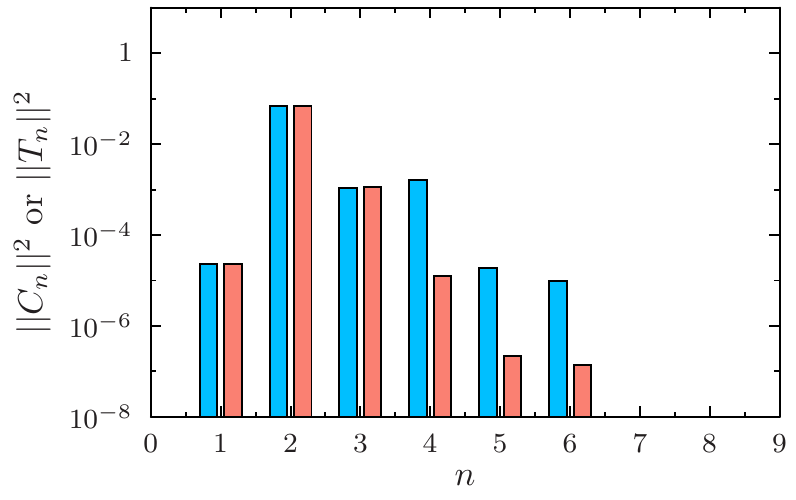}} \subfloat[$r_{\text{OH}}^{0}$, 1M\label{fig:h2o-eq-cc-pvtz-1M}]{\includegraphics[width=0.25\textwidth]{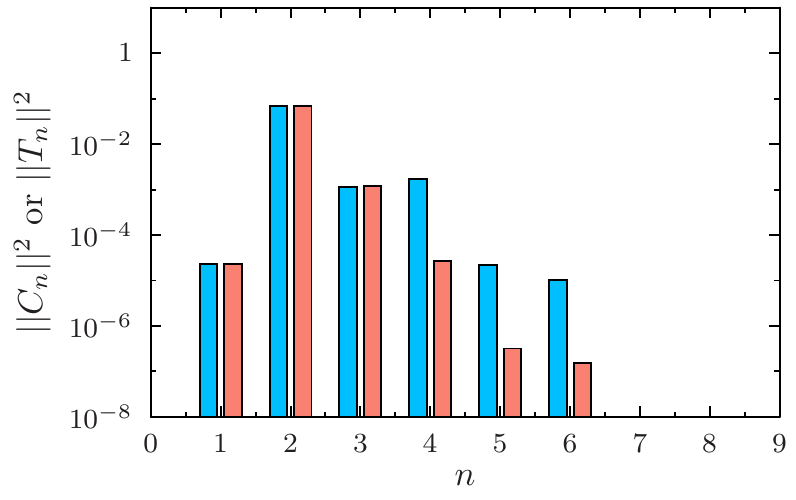}}

  \subfloat[$2r_{\text{OH}}^{0}$, first 1k\label{fig:h2o-20-cc-pvtz-1k}]{\includegraphics[width=0.25\textwidth]{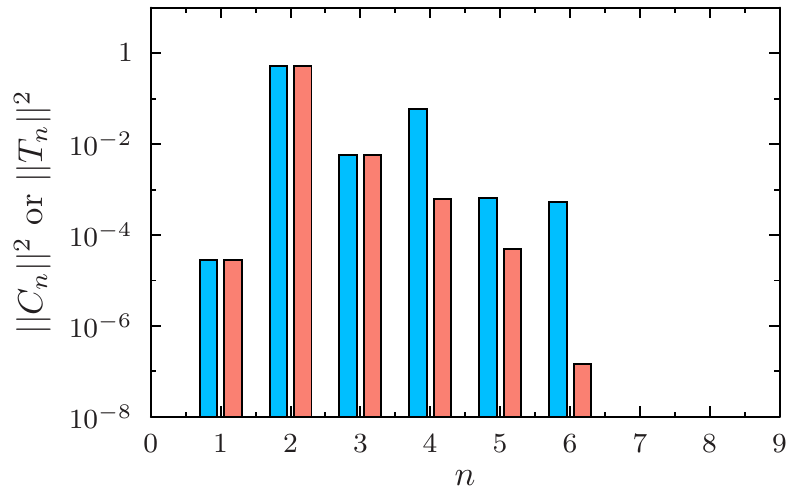}} \subfloat[$2r_{\text{OH}}^{0}$, first 10k\label{fig:h2o-20-cc-pvtz-10k}]{\includegraphics[width=0.25\textwidth]{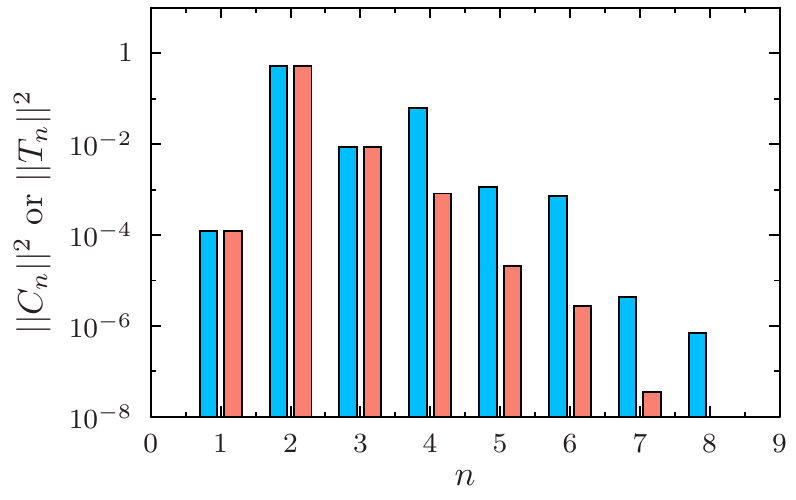}} \subfloat[$2r_{\text{OH}}^{0}$, first 100k\label{fig:h2o-20-cc-pvtz-100k}]{\includegraphics[width=0.25\textwidth]{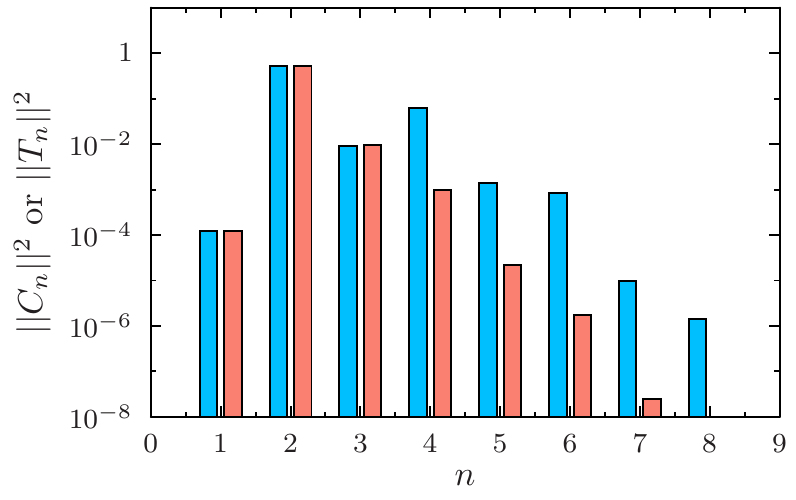}} \subfloat[$2r_{\text{OH}}^{0}$, 1M\label{fig:h2o-20-cc-pvtz-1M}]{\includegraphics[width=0.25\textwidth]{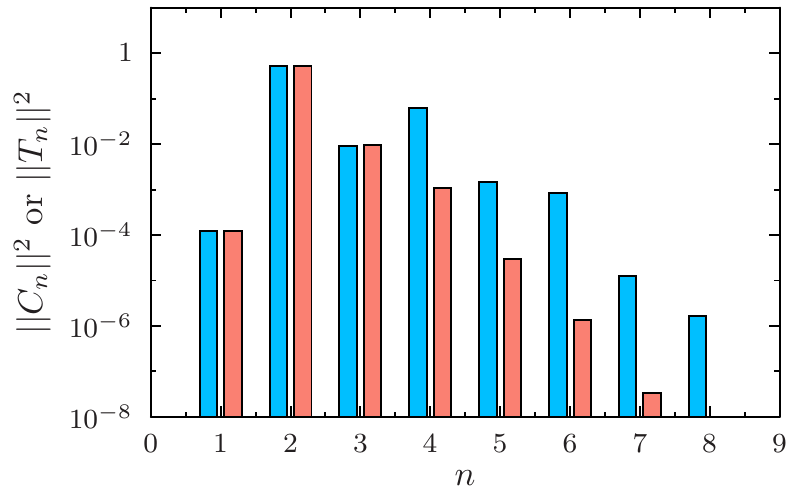}}

  \subfloat[$3r_{\text{OH}}^{0}$, first 1k\label{fig:h2o-30-cc-pvtz-1k}]{\includegraphics[width=0.25\textwidth]{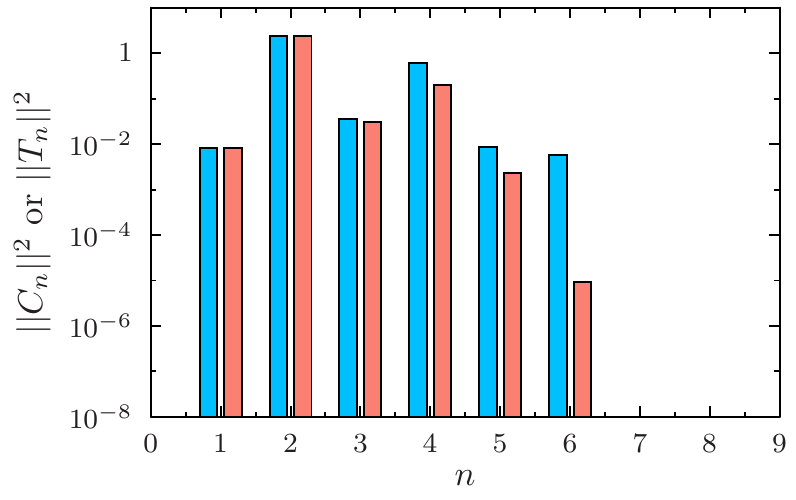}} \subfloat[$3r_{\text{OH}}^{0}$, first 10k\label{fig:h2o-30-cc-pvtz-10k}]{\includegraphics[width=0.25\textwidth]{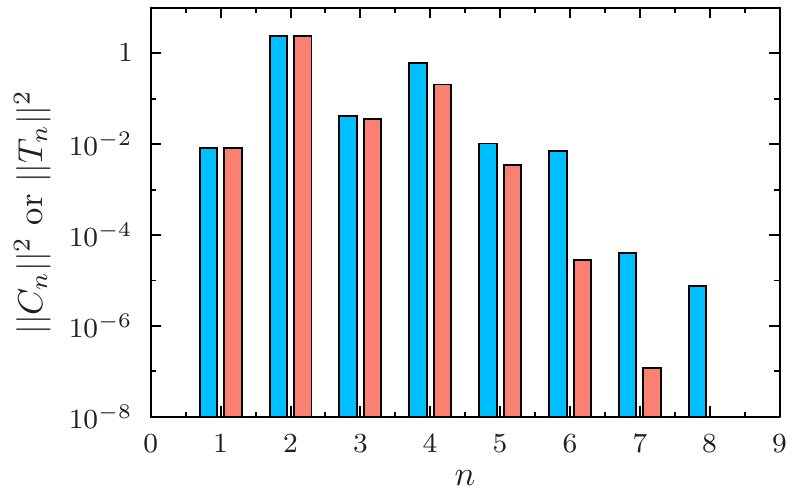}} \subfloat[$3r_{\text{OH}}^{0}$, first 100k\label{fig:h2o-30-cc-pvtz-100k}]{\includegraphics[width=0.25\textwidth]{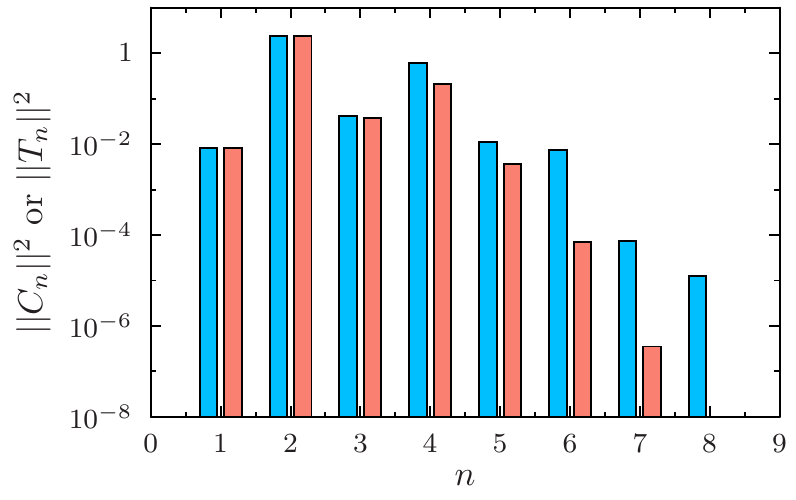}} \subfloat[$3r_{\text{OH}}^{0}$, 1M\label{fig:h2o-30-cc-pvtz-1M}]{\includegraphics[width=0.25\textwidth]{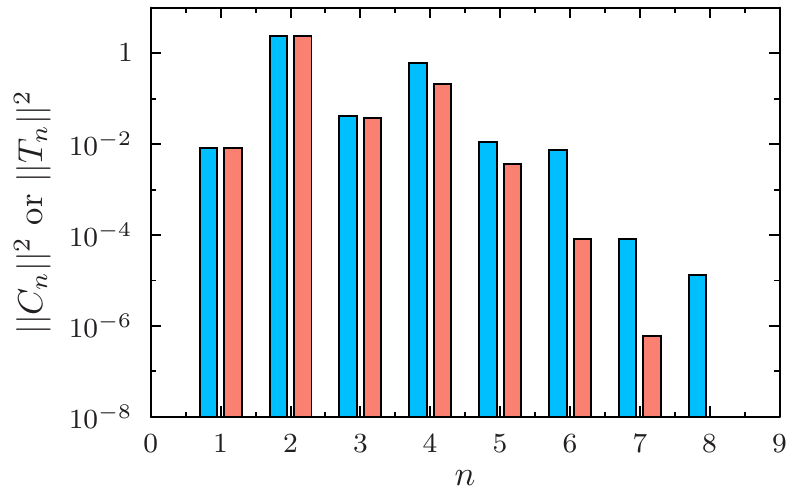}}

  \caption{CI (blue) and CC (red) coefficients for an ASCI wave
    function of water in cc-pVTZ basis at various OH bond lengths
    $r_{\text{OH}}$, plots with the first 1k, 10k, 100k, and all 1M
    determinants, increasing from the left. Note logarithmic
    scale.\label{fig:Water-ccpvtz}}
\end{figure*}

Repeating the calculation in the cc-pVTZ basis\cite{Dunning1989}
yields (10e,58o) i.e. a Hilbert space of $2.1\times10^{13}$
determinants, which is intractable with the conventional FCI
algorithm; however, the calculation is a simple one when adaptive
methods are employed. The results for a 1M determinant ASCI wave function in a
natural orbital basis are shown in \figref{Water-ccpvtz}.  As before,
the decomposition has converged with 100k determinants, and the plots look
much like the 6-31G ones. The largest differences between the exact
6-31G FCI wave function and the cc-pVTZ ASCI wave function are seen at
the equilibrium geometry, where the ASCI wave function appears to lack
highly excited determinants.  However, these have very small norm even
in 6-31G and are likely due to dynamic correlation, which the ASCI
method can capture with a perturbation theory
correction\cite{Tubman2016} that has not been used here.

\subsection{Carbon dimer\label{sub:CarbonDimer}}

\begin{figure*}
  \subfloat[DZ HF, first 1k]{\includegraphics[width=0.33\textwidth]{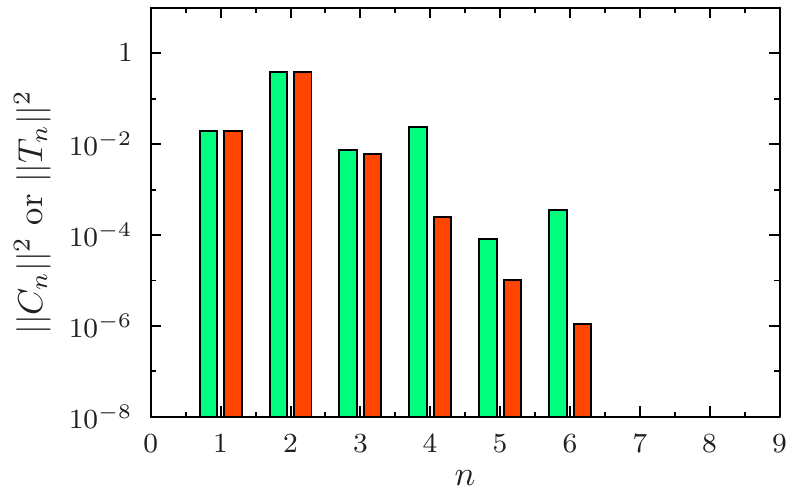}} \subfloat[DZ HF, first 10k]{\includegraphics[width=0.33\textwidth]{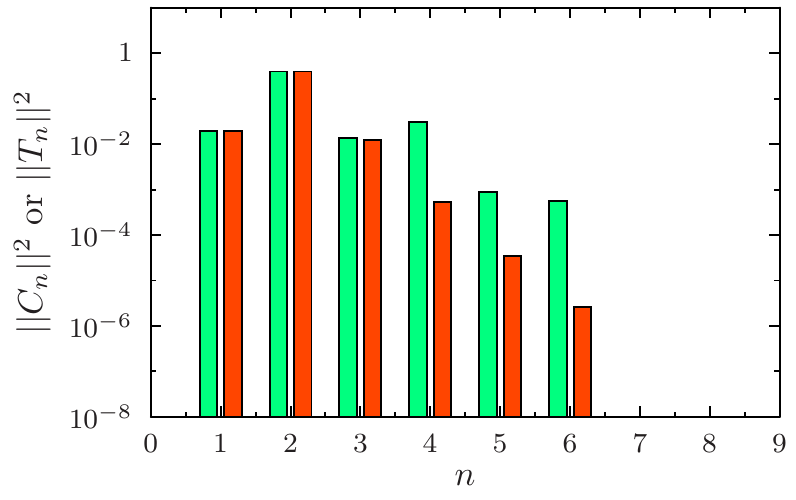}} \subfloat[DZ HF, first 1M]{\includegraphics[width=0.33\textwidth]{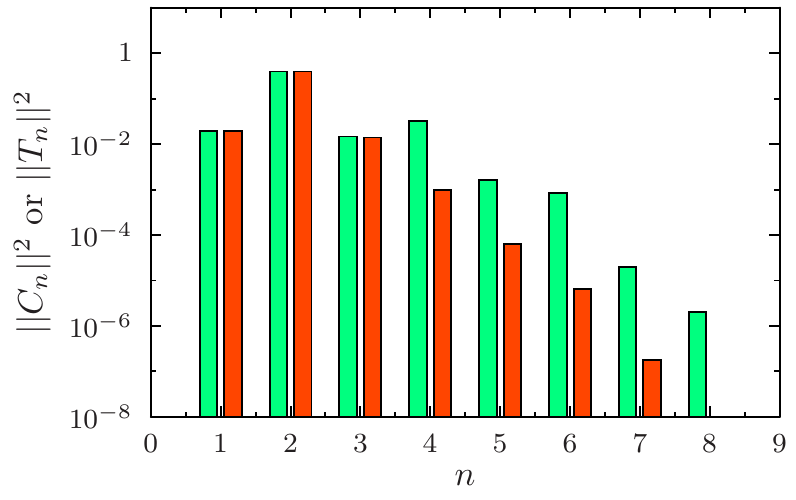}}

  \subfloat[TZ HF, first 1k]{\includegraphics[width=0.33\textwidth]{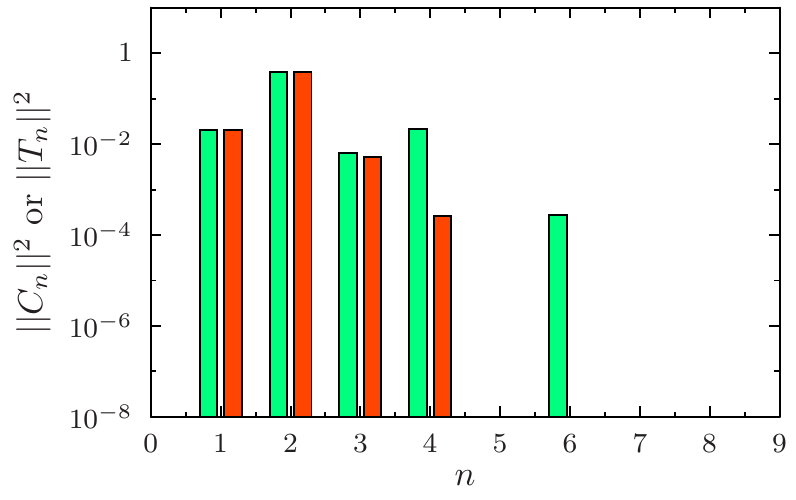}} \subfloat[TZ HF, first 10k]{\includegraphics[width=0.33\textwidth]{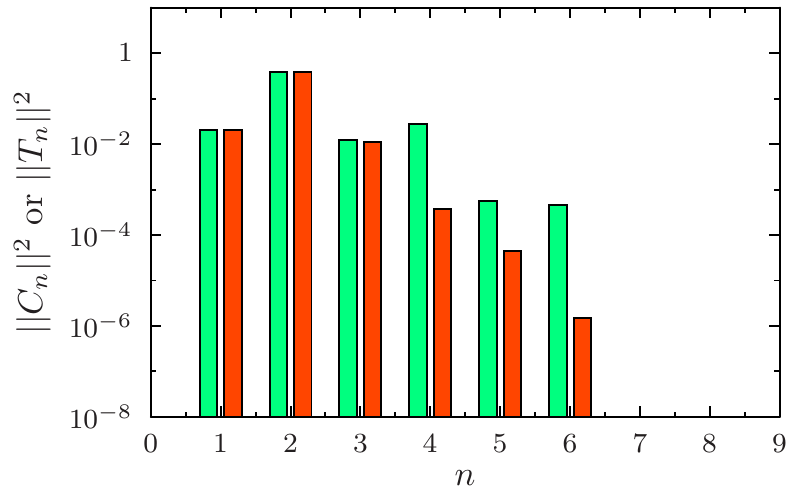}} \subfloat[TZ HF, first 1M]{\includegraphics[width=0.33\textwidth]{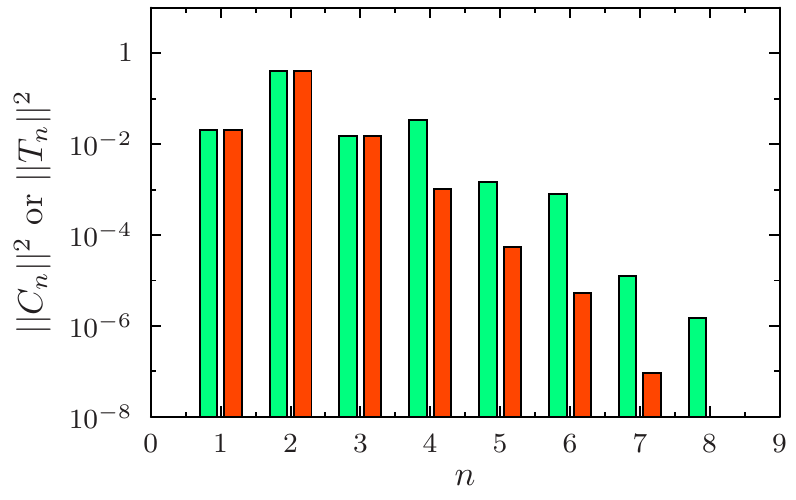}}

  \subfloat[TZ NO, first 1k]{\includegraphics[width=0.33\textwidth]{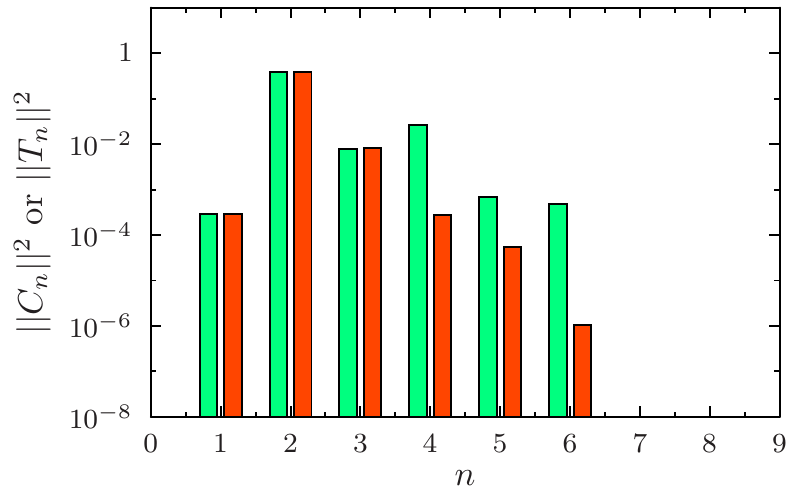}} \subfloat[TZ NO, first 10k]{\includegraphics[width=0.33\textwidth]{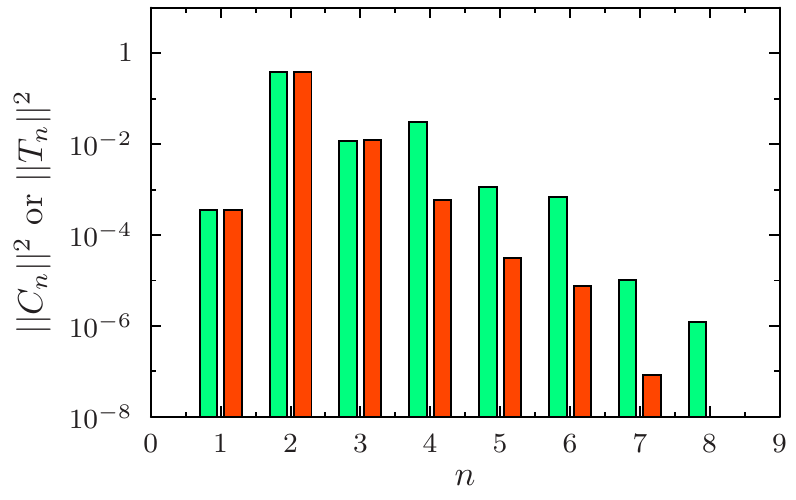}} \subfloat[TZ NO, first 1M]{\includegraphics[width=0.33\textwidth]{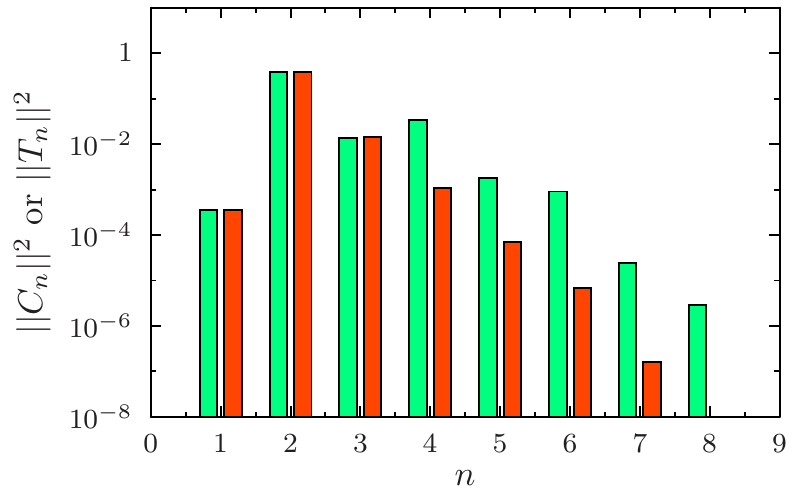}}

\caption{CI (green) and CC (orange) coefficients for double-$\zeta$
  (DZ) and triple-$\zeta$ (TZ) ASCI wave functions of the carbon dimer
  at $R=2.2$ Å with Hartree--Fock (HF) or natural orbitals (NOs),
  plots with 1k, 10k, and 1M determinants, increasing from the
  left. Note logarithmic scale.\label{fig:Carbon-dimer-at}}
\end{figure*}

The carbon dimer is a well-known multiconfigurational system which has
been studied by multiple authors.\cite{Leininger2000, Abrams2004,
  Booth2011, Tubman2016} In \figref{Carbon-dimer-at}, we show the
decompositions for the carbon dimer at a stretched geometry,
$r_{\text{CC}}=2.2$ Å, for three kinds of orbitals: double-$\zeta$ HF,
triple-$\zeta$ HF, and triple-$\zeta$ natural orbitals (NOs), with a
varying number of determinants in the wave function. The double-$\zeta$
calculations use the cc-pVDZ basis, yielding (12e,28o) for which
$\dim\mathcal{H}\approx1.4\times10^{11}$, whereas the
triple-$\zeta$ use cc-pVTZ yielding (12e,60o) and
$\dim\mathcal{H}\approx2.5\times10^{15}$.  The ASCI wave functions contain 25M
determinants for the double-$\zeta$ calculation, 7.1M determinants for the
triple-$\zeta$ calculation with HF orbitals, and 10.2M determinants for the
triple-$\zeta$ calculation with NOs.

The resulting decompositions are shown in \figref{Carbon-dimer-at}.
Based on the decomposition, \ce{C2} at $r_{\text{CC}}=2.2$ Å does not
appear to be very strongly correlated, as the $T$-amplitudes bear the
hallmark of dynamic correlation, i.e. a rapid convergence to zero with
increasing rank. Judging from the decomposition, the role of connected
octuple excitations in the 12-electron system is negligible.

As in the previous case of water, the cluster decomposition is seen to
converge rapidly with the increasing size of the CI wave function.
The triple-$\zeta$ calculations are an interesting comparison, as
according to Thouless' theorem a rotation of the orbitals is
tantamount to $T_{1}$.\cite{Thouless1960} Indeed, $T_{1}$ is much
smaller for the NOs than for the HF orbitals, as is evident from
\figref{Carbon-dimer-at}.  While in theory $T_{n}$ for $n>1$ should
match between the HF and NO calculations, this is no longer true when
the wave functions are truncated, but the differences still remain small.

\subsection{Polyacenes\label{sub:Polyacenes}}

\begin{figure*}
  \subfloat[2acene, first 100]{\includegraphics[width=0.25\textwidth]{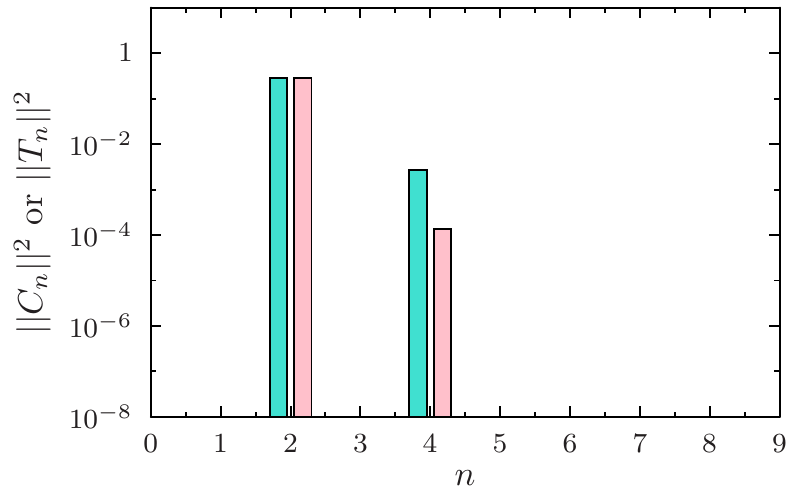}} \subfloat[2acene, first 1k]{\includegraphics[width=0.25\textwidth]{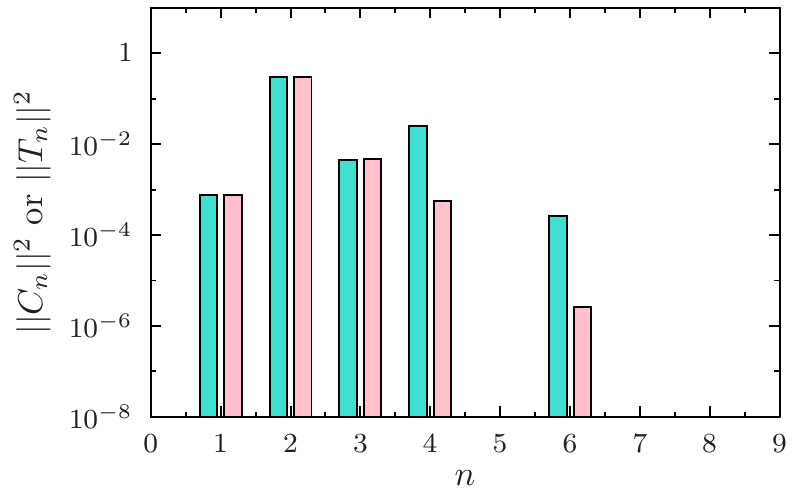}} \subfloat[2acene, first 10k]{\includegraphics[width=0.25\textwidth]{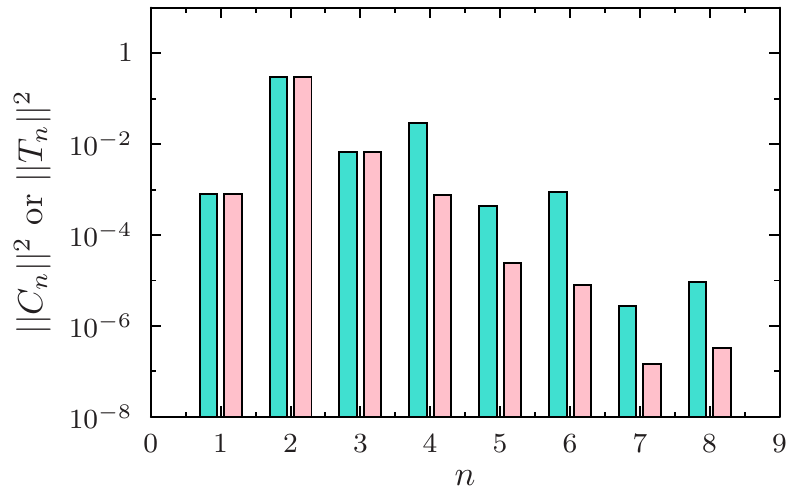}} \subfloat[2acene, 12k]{\includegraphics[width=0.25\textwidth]{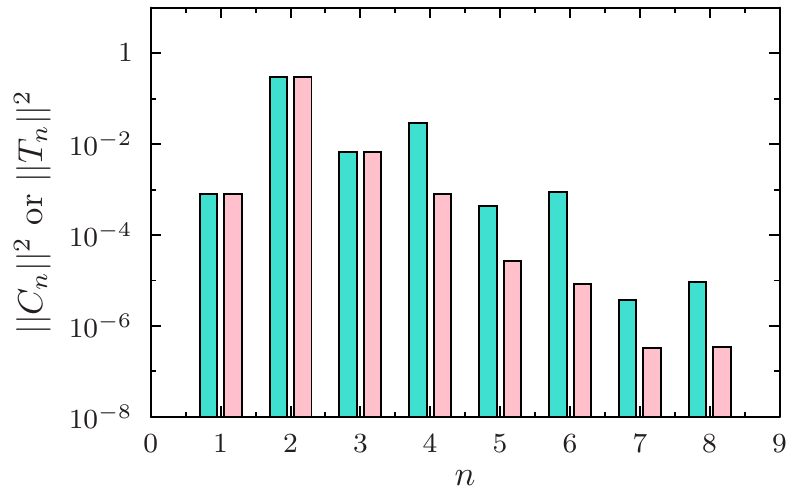}}

  \subfloat[3acene, first 1k]{\includegraphics[width=0.25\textwidth]{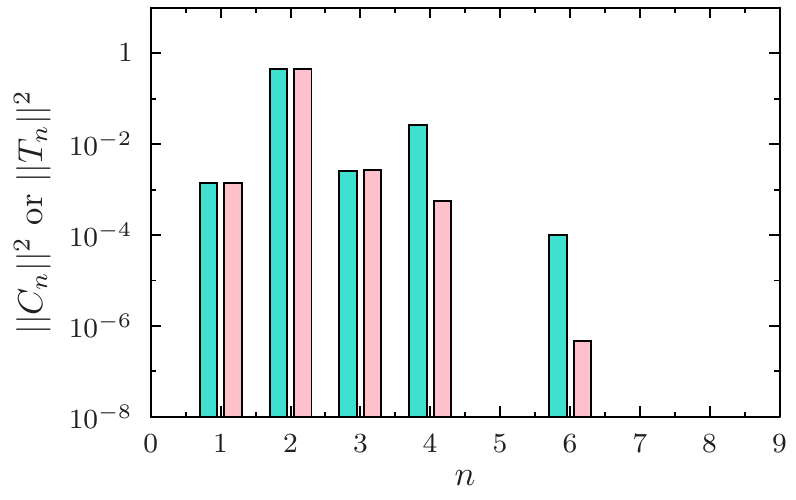}} \subfloat[3acene, first 10k]{\includegraphics[width=0.25\textwidth]{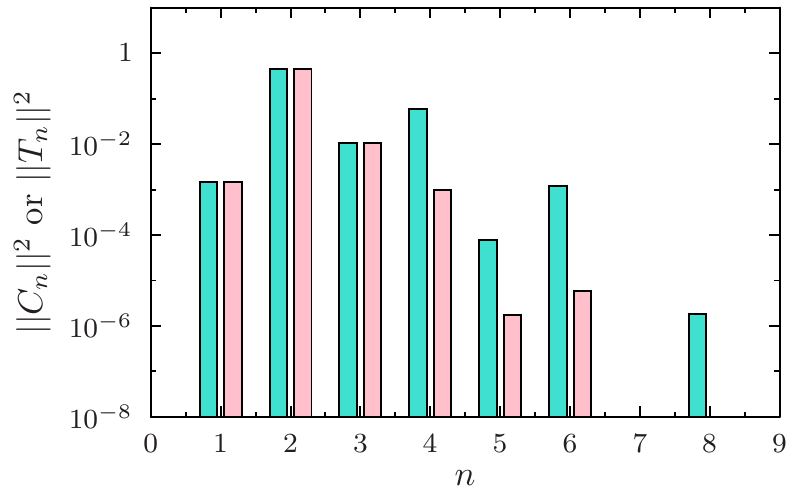}} \subfloat[3acene, first 100k]{\includegraphics[width=0.25\textwidth]{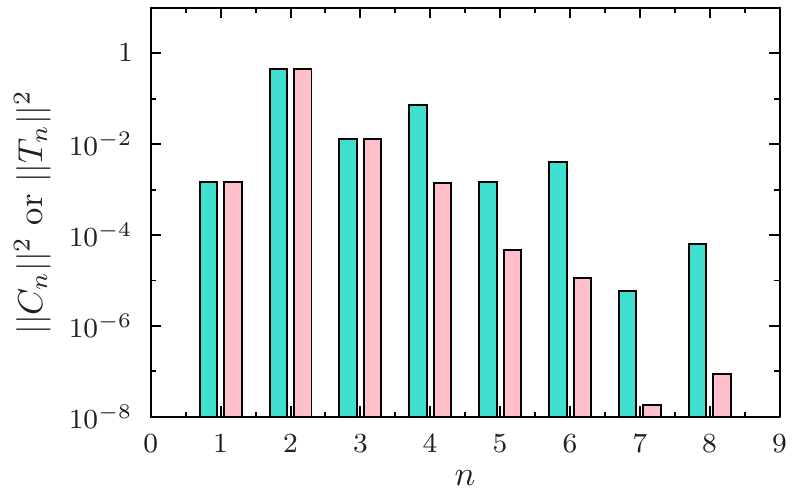}} \subfloat[3acene, 1M]{\includegraphics[width=0.25\textwidth]{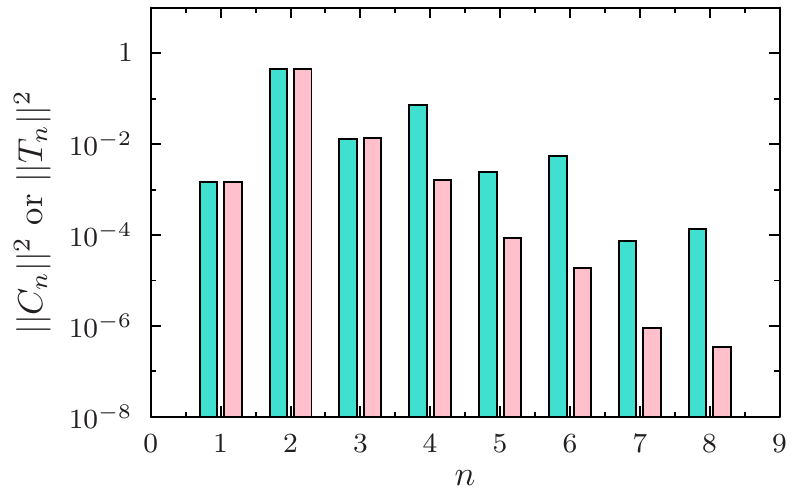}}

  \subfloat[4acene, first 10k]{\includegraphics[width=0.25\textwidth]{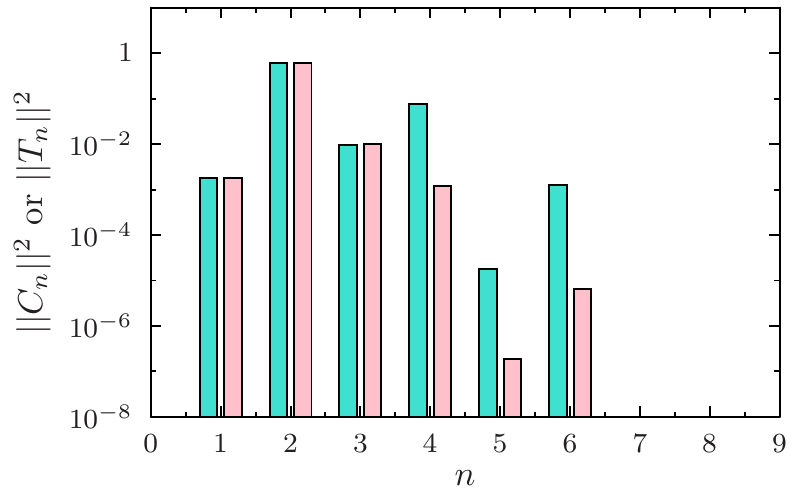}} \subfloat[4acene, first 100k]{\includegraphics[width=0.25\textwidth]{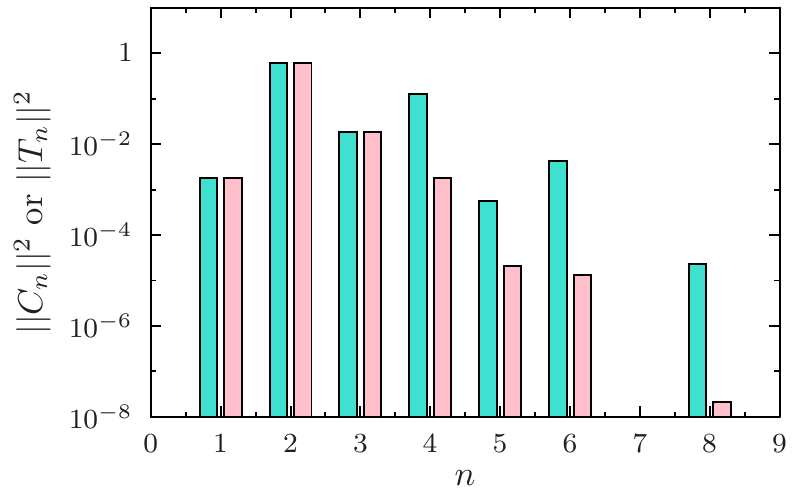}} \subfloat[4acene, first 1M]{\includegraphics[width=0.25\textwidth]{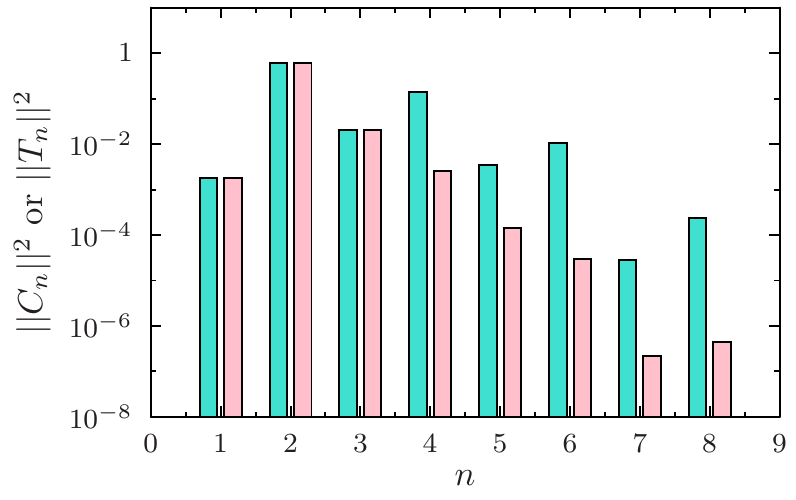}} \subfloat[4acene, first 4M]{\includegraphics[width=0.25\textwidth]{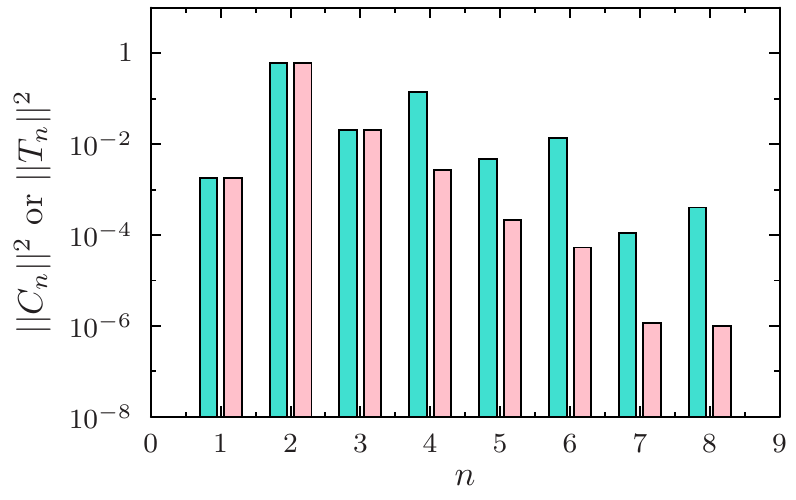}}

  \subfloat[6acene, first 10k]{\includegraphics[width=0.25\textwidth]{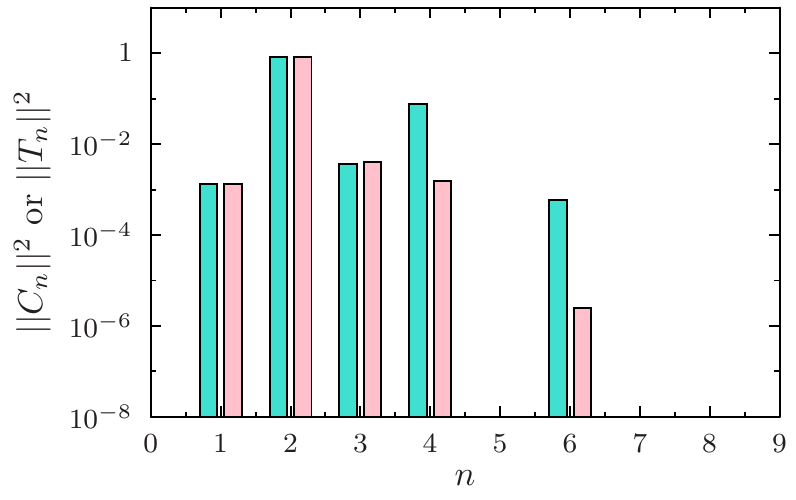}} \subfloat[6acene, first 100k]{\includegraphics[width=0.25\textwidth]{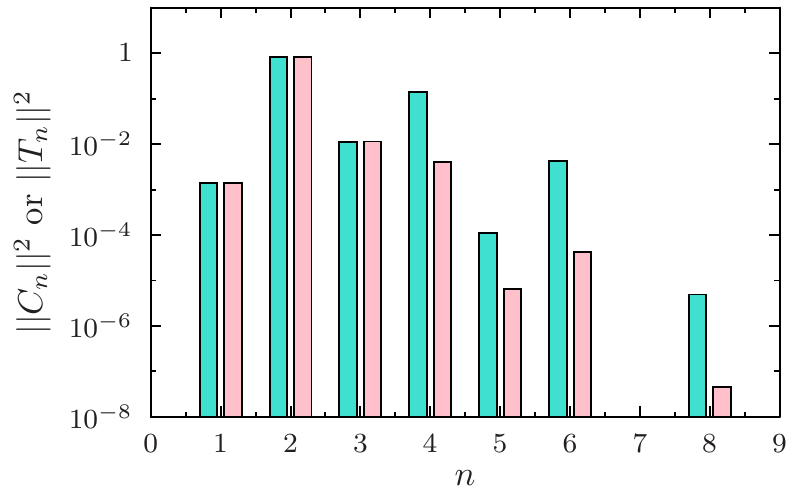}} \subfloat[6acene, first 1M]{\includegraphics[width=0.25\textwidth]{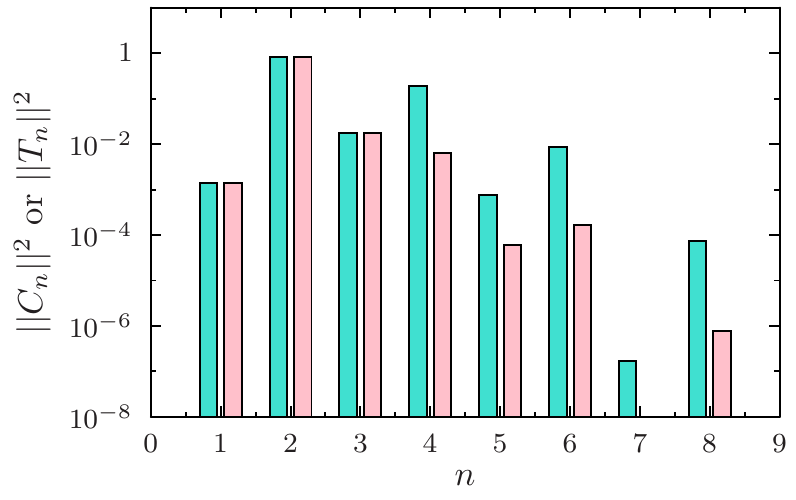}} \subfloat[6acene, first 4M]{\includegraphics[width=0.25\textwidth]{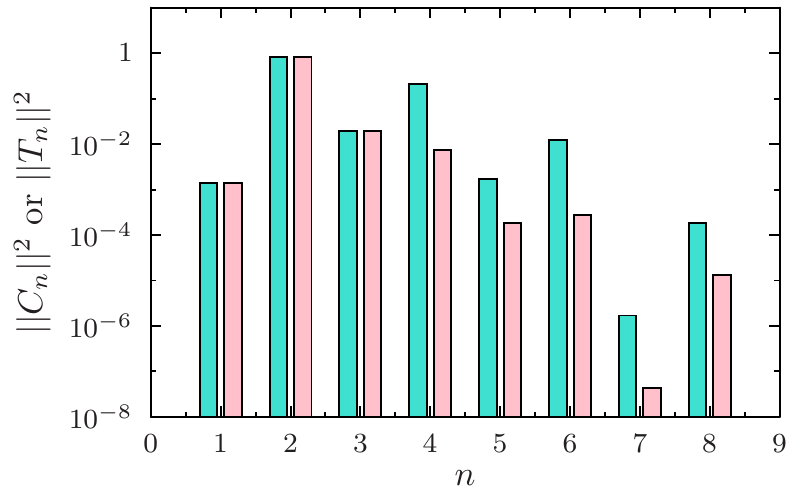}}

\caption{CI (turquoise) and CC (pink) coefficients for STO-3G
  $\pi$-space ASCI wave functions in polyacenes. \label{fig:acene-pi}}
\end{figure*}

Polyacenes are thought to exhibit strong correlation in the $\pi$
space, based on density matrix renormalization group calculations in
the STO-3G basis set;\cite{Hachmann2007} for more discussion, see
\citeref{Lehtola2017} and references therein. Some of us have recently
argued that connected quintuple and hextuple excitations have only a small role
in the strong correlation effects, even for the largest
acenes.\cite{Lehtola2017} We demonstrate this in the STO-3G basis with
the (10e,10o), (14e,14o), (18e,18o) and (26e,26o) ASCI wave functions
of 2acene, 3acene, 4acene and 6acene, respectively, shown in
\figref{acene-pi}, using geometries from \citeref{Hachmann2007}. For
2acene, a 12k determinant ASCI wave function was used, which is almost
FCI for this system. A 1M determinant ASCI wave function was used for
3acene, whereas the 4acene and 6acene wave functions contained over 6M
determinants.

Comparing the cluster decompositions for 2acene, 3acene and 4acene
shows that they are strikingly similar. Connected doubles are by far
the most important amplitudes, followed by connected triples. Singles
and quadruples are equally important, with quintuples and higher
excitations being an order of magnitude less important, displaying the
signature of dynamic correlation as was seen above in the case of
water.

Due to its considerably larger Hilbert space, the 6acene decomposition
appears still far from convergence at 4M determinants. But, in all the wave
functions the convergence behavior of the cluster decomposition is
apparent: the excitations converge roughly order of increasing rank.
Comparison to the non-converged truncations of the smaller acenes
again reveals striking similarities between the wave functions.

These findings thus appear to fully confirm the speculation in
\citeref{Lehtola2017}: connected quadruple excitations should suffice
for the treatment of static correlation in the acenes.

\subsection{Chromium dimer\label{sub:ChromiumDimer}}

The chromium dimer is a challenging system which is not well described
by even CCSDTQ, as a 14 m$E_{h}$ difference has been found in the
total energy in a (24e,30o) active space at an interatomic distance of
$R=1.7$ Å.\cite{Kurashige2009} Here, we study \ce{Cr2} in a Karlsruhe
split-valence (SV) basis set,\cite{Schafer1992} with (48e,42o), giving
a size of the Hilbert space of $1.3\times10^{23}$.

\begin{figure*}
  \subfloat[$R=1.6$ Å, first 10k]{\includegraphics[width=0.25\textwidth]{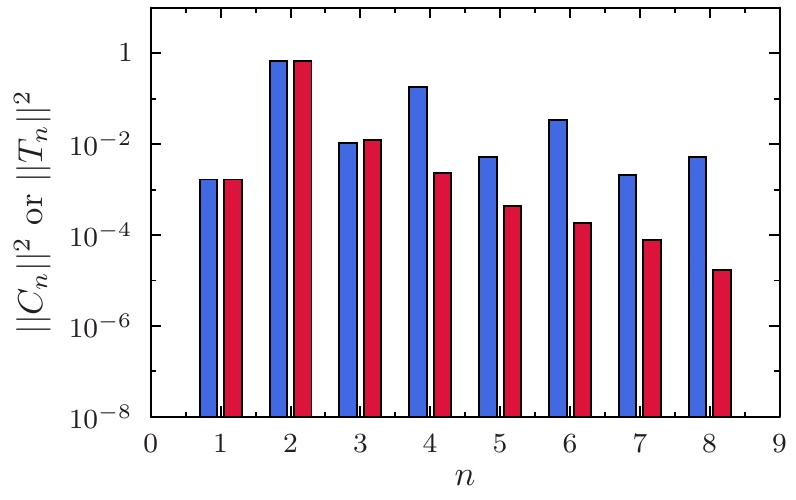}} \subfloat[$R=1.6$ Å, first 100k]{\includegraphics[width=0.25\textwidth]{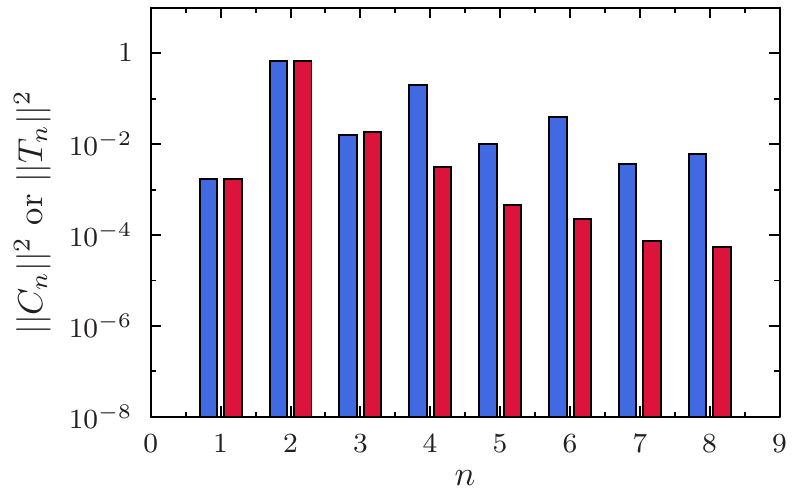}} \subfloat[$R=1.6$ Å, first 1M]{\includegraphics[width=0.25\textwidth]{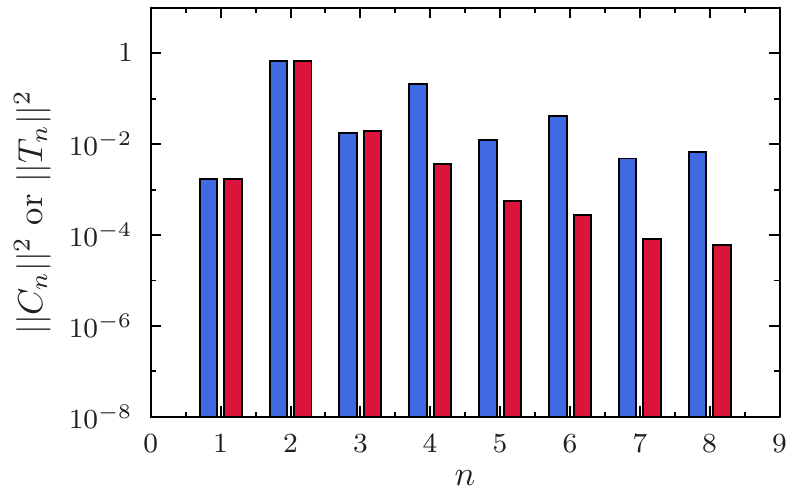}} \subfloat[$R=1.6$ Å, first 4M]{\includegraphics[width=0.25\textwidth]{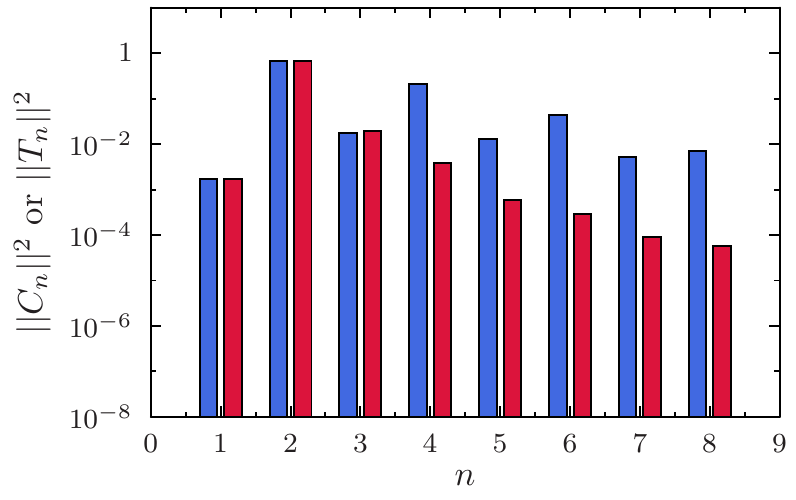}}

  \subfloat[$R=1.8$ Å, first 10k]{\includegraphics[width=0.25\textwidth]{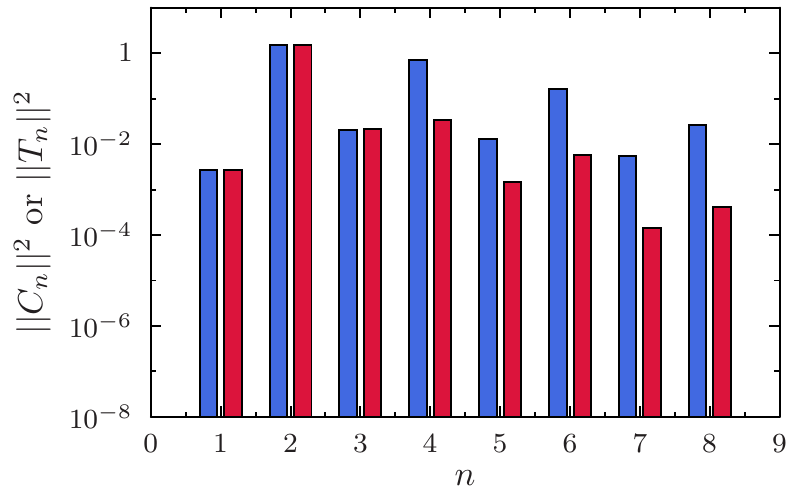}} \subfloat[$R=1.8$ Å, first 100k]{\includegraphics[width=0.25\textwidth]{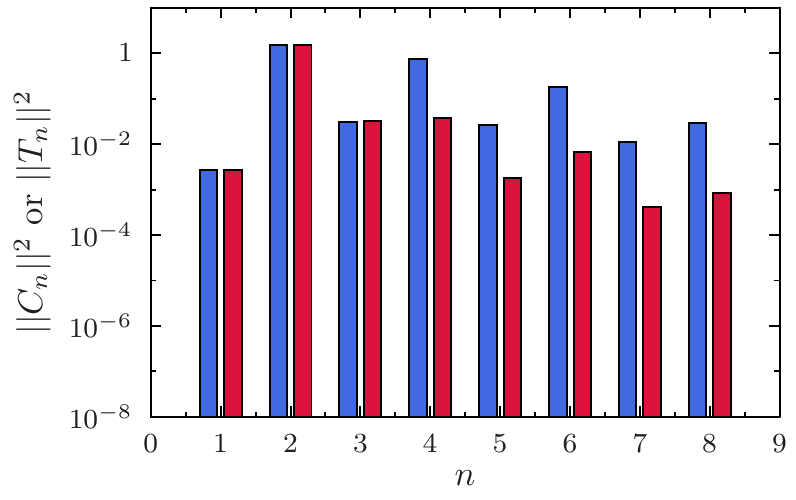}} \subfloat[$R=1.8$ Å, first 1M]{\includegraphics[width=0.25\textwidth]{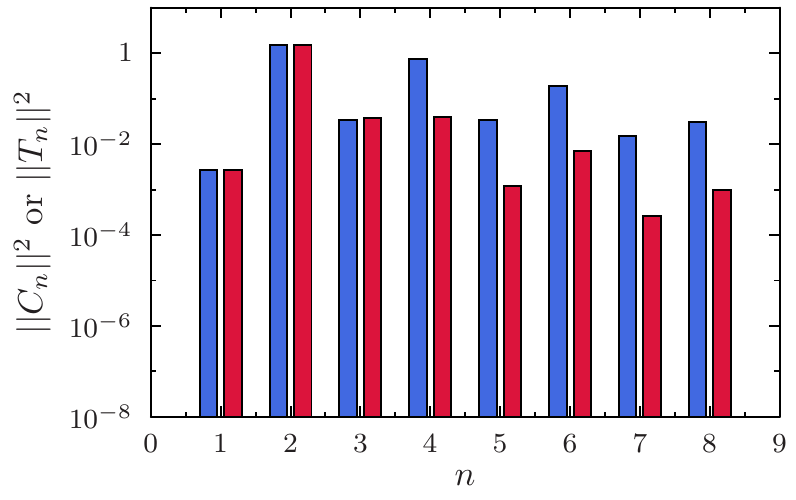}} \subfloat[$R=1.8$ Å, first 4M]{\includegraphics[width=0.25\textwidth]{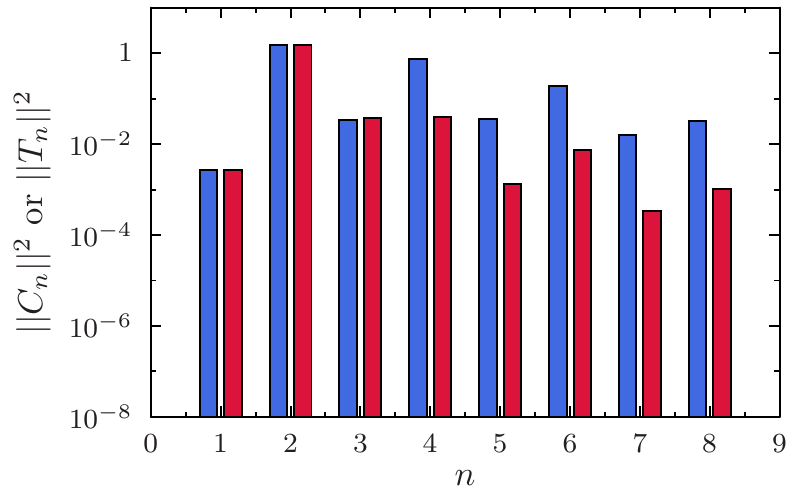}}
  
  \subfloat[$R=2.0$ Å, first 10k]{\includegraphics[width=0.25\textwidth]{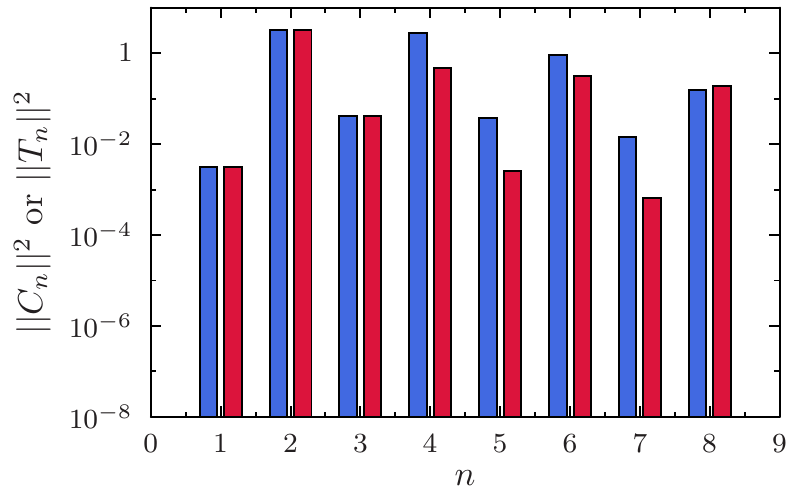}} \subfloat[$R=2.0$ Å, first 100k]{\includegraphics[width=0.25\textwidth]{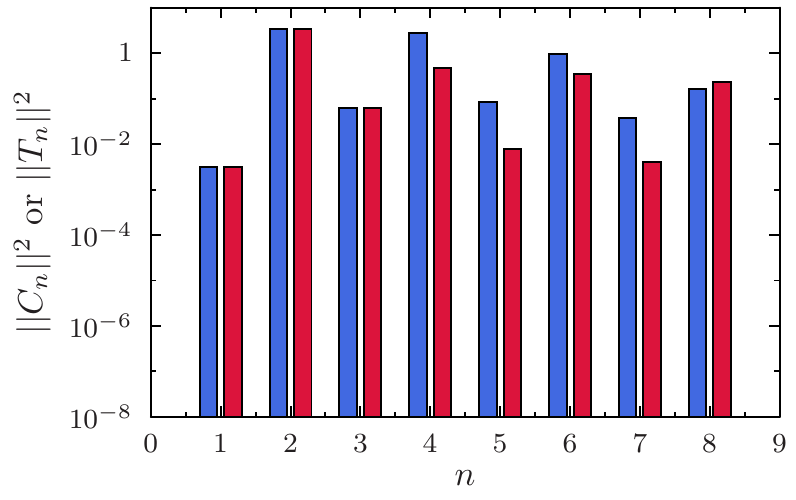}} \subfloat[$R=2.0$ Å, first 1M]{\includegraphics[width=0.25\textwidth]{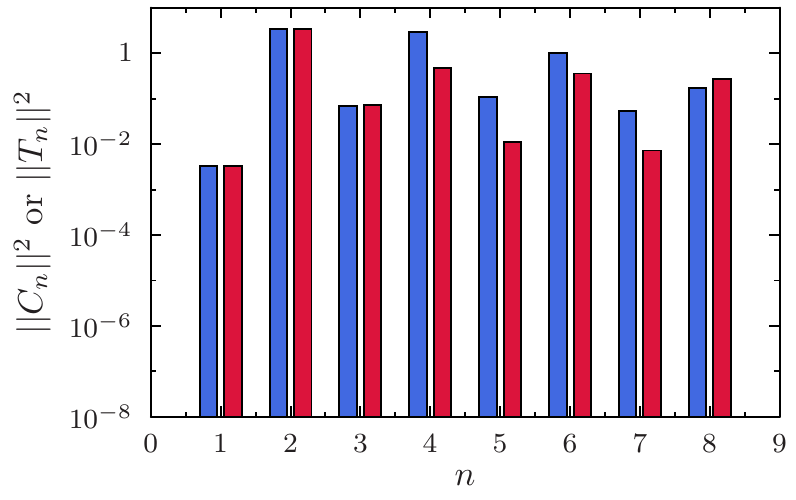}} \subfloat[$R=2.0$ Å, first 4M]{\includegraphics[width=0.25\textwidth]{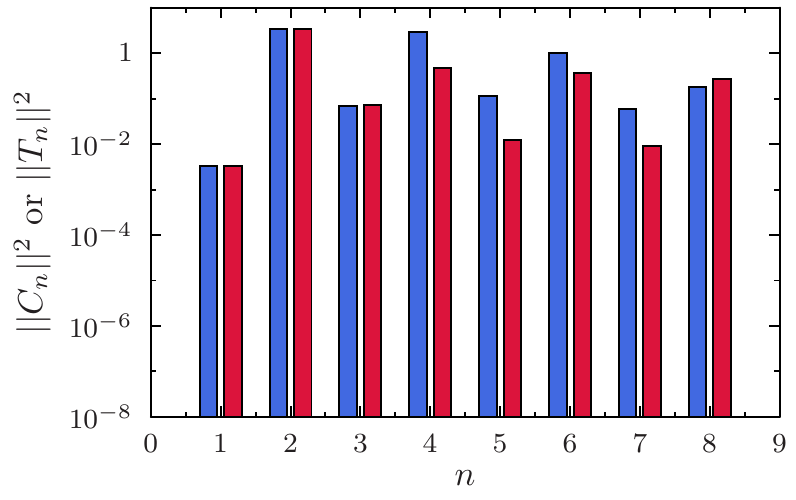}}

\caption{CI (blue) and CC (red) coefficients for chromium dimer with
  first 10k, 100k, 1M and 4M determinants out of a $\geq$4.8M
  determinant ASCI wave function . Note logarithmic
  scale.\label{fig:cr2-plots}}
\end{figure*}

The decompositions at $R=1.6$ Å to $R=2.0$ Å based on a natural
orbital reference are shown in \figref{cr2-plots}. It is apparent that
the strong correlation effects increase significantly in magnitude
when the atoms are pulled apart. Furthermore, compared to all the
other systems in the present study, the speed of convergence of the
cluster expansion is alarmingly slow even around $R=1.6$ Å. The
weights of the connected excitations decay only gradually. The
inclusion of all $\left\Vert T_{n}\right\Vert \lesssim10^{-4}$ would
require the description of connected hextuples at $R=1.6$ Å, while
already at $R=1.8$ Å connected octuples become significant. For
distances $R>1.6$ Å the decomposition does not appear converged,
suggesting that approaches based on spin-restricted single-reference
CC may be intractable for many multireference problems in transition
metal chemistry.

\section{Summary and Discussion\label{sec:Summary-and-Discussion}}

We have presented an implementation of the cluster decomposition
technique, which can be used to translate full configuration
interaction (FCI) wave functions into full coupled-cluster (CC)
expansions. The decomposition is also applicable to FCI within an
active space, i.e. for interpreting CASSCF calculations, as well as to
selected CI wave functions such as the ones produced by the ASCI
method.\cite{Tubman2016} The decomposition is based on an iterative
algorithm which subtracts the contributions from disconnected
excitations (independent products of lower-rank excitations) from a
given CI expansion coefficient, forming the CC expansion one rank at a
time.  The complexity of the recursion relations grows rapidly with
increasing rank, due to which our implementation only supports the
conversion up to octuple excitations. Due to their sheer number, the
rate-determining steps in both the generation and application of the
cluster expansion have to do with terms involving products of single
excitations. The decomposition could be pushed much further by
performing the CI calculations with Brueckner
orbitals,\cite{Brueckner1955} which make $T_{1}$ vanish.

The interest in the decomposition stems from the recent emergence of
stochastic and adaptive FCI wave functions, which have made
calculations possible on systems of unforeseen sizes. The technique
which yields information on the connected excitations in a system is
useful in multiple aspects. First, it can be used to analyze which
level of CC theory is necessary to qualitatively describe a given
system. Second, it can serve as a measure of the convergence of the
adaptive FCI wave function in cases where it is known that no
connected excitations should emerge beyond a certain rank (for
example, $C_{n>2}\neq0$ but $T_{n>2}=0$ for a system of
non-interacting helium atoms).

We have studied the convergence of the cluster decomposition based on
FCI as well as ASCI wave functions. For each system, the convergence
with respect to the wave function length was performed by performing
calculations with various truncations of the fixed input wave
function. Although we have introduced a sparse approximation for
carrying out the procedure, the decompositions have been found to
converge rapidly with increasing wave function length, highlighting
the usefulness of our approach.

While the results of the present work are for spin-restricted
orbitals, we believe that the results for calculations with
spin-unrestricted orbitals would be similar. Namely, spin
unrestriction would only decrease $\hat{T}_1$ -- which is relatively
small in all the calculations of the present manuscript -- and
increase the weight of the reference determinant -- which is fairly
large in all the calculations due to the use of suitable orbitals for
each problem.

As is well known, calculations based on the coupled-cluster method
converge rapidly with respect to excitation rank used in the
calculation, which has also been used recently to construct blazing
fast approximate CASSCF approaches.\cite{Parkhill2009, Parkhill2010,
  Parkhill2010b, Parkhill2011, Lehtola2016b, Lehtola2017} However, the
speed of convergence of coupled-cluster theory may still be too low to
be able to cost-efficiently treat challenging strong correlation
problems. Our results on \ce{Cr2} lead us to believe that
single-reference CC approaches cannot be faithfully applied on
challenging problems in transition metal chemistry, leaving room for
alternative approaches.

\section*{Acknowledgments}
This work was supported through the Scientific Discovery through
Advanced Computing (SciDAC) program funded by the U.S. Department of
Energy, Office of Science, Advanced Scientific Computing Research and
Basic Energy Sciences, and by the Director, Office of Basic Energy
Sciences, Chemical Sciences, Geoscience and Biosciences Division of
the US Department of Energy under Contract
No. DE-AC02-05CH11231. Computational resources provided by the Extreme
Science and Engineering Discovery Environment (XSEDE), which is
supported by the National Science Foundation Grant No. OCI-1053575,
are gratefully acknowledged.

\section*{Appendix}

The format used by the \textsc{ClusterDec} program\cite{clusterdec} is
the following:

$N_{\text{dets}}$ $N_{\text{orb}}$ $N_{\alpha}$ $N_{\beta}$

$\text{coeff}_{1}$ $\text{bitstring}_{1}$

$\vdots$

$\text{coeff}_{N_{\text{dets}}}$ $\text{bitstring}_{N_{\text{dets}}}$

\noindent where $N_{\text{dets}}$, $N_{\text{orb}}$, $N_{\alpha}$, and
$N_{\beta}$ denote the number of determinants, spatial molecular
orbitals, and alpha and beta electrons, respectively. The rest of the
file contains the $N_{\text{dets}}$ CI coefficient and determinant
string entries.

To run an analysis, the user first calculates a wave function using
e.g. FCI or ASCI, and feeds it to the program. The syntax of the
program is
\begin{verbatim}
  clusterdec wf rank ndets
\end{verbatim}
The wave function contained in file \texttt{wf} is sorted in
decreasing absolute amplitude, is truncated to \texttt{ndets}
determinants if necessary, after which the decomposition is calculated
up to \texttt{rank}.

\end{document}